\documentclass[%
 aip,
rsi,%
 amsmath,amssymb,
 floatfix
reprint,%
]{revtex4-1}

\usepackage{graphicx}
\usepackage{dcolumn}
\usepackage{bm}
\usepackage[colorlinks=true,linkcolor=blue]{hyperref}
\usepackage[version=4]{mhchem}

\usepackage{color}
\usepackage[dvipsnames]{xcolor}
\usepackage{multirow}
\usepackage{rotating}
\usepackage{threeparttable}
\newcommand{\dirac}{\textsc{Dirac}}

\begin{document}

\preprint{AIP/123-QED}

\title{Beyond the electric-dipole approximation in simulations
	of X-ray absorption spectroscopy: Lessons from relativistic theory}

\author{Nanna Holmgaard List}
 \email{nhlist@stanford.edu}
 \affiliation{Department of Chemistry and the PULSE Institute, Stanford
University, Stanford, California 94305, USA.}
\affiliation{SLAC National Accelerator Laboratory, Menlo Park, California
94025, USA.}

\author{Timoth\'e Romain L\'eo Melin}%
\affiliation{Laboratoire de Chimie et Physique Quantique, UMR 5626 CNRS
--- Universit\'e Toulouse III-Paul Sabatier, 118 route de Narbonne,
F-31062 Toulouse, France}
\affiliation{Present address: Department of Chemistry, Michigan State University, 
			East Lansing, Michigan 48824, USA.}

\author{Martin van Horn}%
\affiliation{Laboratoire de Chimie et Physique Quantique, UMR 5626 CNRS
	--- Universit\'e Toulouse III-Paul Sabatier, 118 route de Narbonne,
	F-31062 Toulouse, France}

\author{Trond Saue}
\email{trond.saue@irsamc.ups-tlse.fr}
\homepage{http://dirac.ups-tlse.fr/saue}
\affiliation{Laboratoire de Chimie et Physique Quantique, UMR 5626 CNRS
--- Universit\'e Toulouse III-Paul Sabatier, 118 route de Narbonne,
F-31062 Toulouse, France}

\date{\today}

\begin{abstract}

  We present three schemes to go beyond the electric-dipole approximation in X-ray absorption spectroscopy calculations within a four-component relativistic framework. The first is based on the full semi-classical light-matter interaction operator, and the two others on a truncated interaction within Coulomb gauge (velocity representation) and multipolar gauge (length representation). We generalize the derivation of multipolar gauge to an arbitrary
  expansion point and show that the potentials corresponding to different expansion point are related by a gauge transformation, provided the expansion is not truncated. This suggests that the observed gauge-origin dependence in multipolar gauge is more than just a finite-basis set effect. 
The simplicity of the relativistic formalism enables \textit{arbitrary-order} implementations of the truncated interactions, with and without rotational averaging, allowing us to test their convergence behavior numerically by comparison to the full formulation. We confirm the
observation that the oscillator strength of the electric-dipole allowed ligand \textit{K}-edge transition of \ce{TiCl4}, when calculated to second order in the
wave vector, become negative, but also show that inclusion of higher-order contributions allows convergence to the result obtained using the full
light-matter interaction. However, at higher energies, the slow convergence of such expansions becomes dramatic and renders such approaches
at best impractical. When going beyond the electric-dipole approximation, we therefore recommend the use of the full light-matter interaction.
\end{abstract}

\pacs{Valid PACS appear here}
\maketitle

\section{Introduction}

The importance of relativistic effects in chemistry is illustrated
by the fact that without relativity gold would have the same color
as silver,\cite{Christensen_PhysRevB1971,Romaniello_JCP2005,Glantschnig_NJP2010}
mercury would not be liquid at room temperature\cite{Calvo_AngChem2013,Steenbergen_JPCL2017}
and your car, if using a lead battery, would not start.\cite{Pekka_PhysRevLett.106.018301}
The present work highlights another aspect of relativity, namely its
essential role in light-matter interactions. 

A semi-classical treatment invoking the electric-dipole (ED) approximation is a common starting point for a theoretical description of light-matter interactions. 
The latter approximation assumes that the spatial extent of the molecular system is small compared to the wavelength of the electromagnetic field, such that the molecule effectively sees a uniform electric field while the magnetic field component is neglected. Formally it corresponds to retaining only the zeroth-order term of an expansion of the interaction operator in orders of the length of the wave vector. While this is often well-justifiable 
for the most commonly used optical laser sources and intensities, the availability of i) high-energy X-ray photons, with wavelengths comparable to the molecular target,\cite{goppert1931elementarakte,List_JCP2015,demekhin2014breakdown} and ii) intense laser sources, creating high-energy electrons strongly influenced by the magnetic component of the Lorentz force,\cite{katsouleas1993comment,reiss1992theoretical,reiss2000dipole} motivates investigations into the effects of going beyond this simplification. Clearly, in either limit, relativistic effects become increasingly important, as the velocity of the electron being probed or driven by the laser field reaches a substantial fraction of the speed of light.

In this work, we focus on going beyond the electric-dipole (BED) approximation in relativistic simulations of near-edge X-ray absorption spectroscopy. While non-dipolar corrections to the total cross sections first enter at second order and are generally quite small ($5-10\%$ for dipole-allowed \textit{K}-edge transitions in the soft X-ray region, reaching up ${\sim}20\%$ in the hard X-ray region\cite{List_JCP2015}), the important \textit{K} pre-edge features may, as is often the case in transition metal complexes, be (near) electric-dipole-forbidden.\cite{shulman1976observations,drager1988multipole,yamamoto2008assignment} 
In general, methods for going BED approximation have been based on multipole expansions of the minimal coupling light-matter interaction operator which, in truncated form, may introduce unphysical gauge-origin dependence into the molecular properties.\cite{george2008time} This is particularly problematic for molecular systems where no natural choice of gauge origin exists.
In a seminal paper, Bernadotte \textit{et al.}~presented an approach for the calculation of origin-independent intensities within the non-relativistic framework, beyond the ED approximation, by truncating the oscillator strength, rather than the interaction operator, in orders of the wave vector.\cite{bernadotte_JCP2012} In the velocity representation, they could demonstrate origin independence of oscillator strengths to arbitrary order, and
could confirm this by calculation to second order. Bernadotte \textit{et al.} furthermore transformed the interaction operator truncated to second order in
the wave vector from its velocity representation to multipolar form (for earlier demonstrations of this transformation see for instance Refs.  \citenum{Barron_Gray_JPA1973,barron2004molecular}). This would imply origin independence of oscillator strengths to arbitrary order also in the length representation, but this was not observed in calculations to second order and attributed to the finite basis approximation. Further complications were
reported by Lestrange \textit{et al.} \cite{lestrange_JCP2015} who found that including the second-order oscillator strength of the ED allowed ligand \textit{K}-edge transition of \ce{TiCl4}
made the total oscillator strength negative. Negative oscillator strengths to second order were also reported by S{\o}rensen \textit{et al.}\cite{Sorensen_MP2016} in metal \textit{K}-edge transitions of \ce{[FeCl4]-}, but only for certain basis sets, which led them to conclude that they were due to incomplete basis sets rather than missing higher-order contributions to the oscillator strength.
In a second paper \cite{sorensen_CPL2017}, where \ce{[FeCl4]-} is revisited, S{\o}rensen \textit{et al.} speculate that the fourth-order electric-octupole-electric-octupole contribution may reverse the sign ``provided that no other higher terms also grows disproportionately large''.

To avoid the above issues, we recently proposed using the \textit{full} semi-classical light-matter interaction operator in the context of linear absorption spectroscopy in the non-relativistic regime.\cite{List_JCP2015} In a Gaussian basis, the necessary integrals over the light-matter interaction operator can be identified as Fourier transforms of overlap distributions, as shown by Lehtola \textit{et al.}~for dynamic structure factors,\cite{lehtola2012erkale} and can be easily evaluated within standard integral schemes, such as McMurchie--Davidson\cite{List_JCP2015} or Gauss--Hermite quadrature.\cite{sorensen2019implementation} In a second paper,\cite{List_MP2017} we presented a mixed analytical-numerical approach to isotropically average oscillator strengths computed with the full light-matter interaction operator.\cite{List_MP2017} This novel approach has been followed up by S{\o}rensen \textit{et al.}\cite{sorensen2019implementation,khamesian2019spectroscopy} Some other works using full light-matter interaction may be mentioned: Kaplan and Markin calculated the photoionization cross section of the \ce{H2} molecule in both a non-relativistic\cite{Kaplan_Dokl1969} and relativistic setting.\cite{Kaplan_JETP1973} More recent work along these lines include Refs.~\citenum{Seabra_JCP2005,demekhin2014breakdown,Brumboiu_JCP2019}. A recent review has been given by
Wang \textit{et al.}\cite{Wang_ChinPhysB2020}

In the following, we present three schemes for computing linear absorption cross sections beyond the ED approximation within a four-component relativistic framework: i) the full semi-classical light-matter interaction as well as two approaches based on truncated interaction either using ii) multipolar gauge (length representation) or iii) Coulomb gauge (velocity representation). The latter may be viewed as an extension of the work by Bernadotte \textit{et al.}\cite{bernadotte_JCP2012}~to the relativistic domain. For all three schemes we present methods for rotational averaging; for the full interaction we use the
mixed analytical-numerical approach already reported for non-relativistic calculations,\cite{List_MP2017} whereas for truncated interaction we have developed a fully analytical approach. 
As will become clear below, in addition to providing a more general framework, the relativistic formalism is more simple than the non-relativistic counterpart and facilitates general, easily programmable expressions. In fact, we have in the \dirac~package \cite{DIRAC19} implemented the two schemes for truncated interaction to \textit{arbitrary order}, with and without rotational averaging,
which allows us to test numerically the convergence behavior of these schemes and compare to the formulation based on the full semi-classical light-matter interaction. 

The paper is organized as follows: In Section \ref{subsec:Coupling-particles-and}, we briefly review the description of 
semi-classical light-matter interactions in both relativistic and non-relativistic frameworks. Section \ref{subsec:Full-light-matter-interaction} 
presents the working expressions for oscillator strengths for the full light-matter interaction operator, followed by a derivation 
of the two different truncated light-matter interaction formulations in Section \ref{subsec:truncated_light_matter}. In 
Section \ref{subsec:rotational_avg}, we describe schemes for obtaining isotropically averaged oscillator strengths in each of the three cases.
In Section \ref{subsec:results_and_discussion} we investigate 
the performance of the three different schemes for going beyond the electric-dipole approximation, before concluding in Section \ref{sec:concl}. 

We also provide three appendices: In appendix \ref{sec:Simulation-of-electronic} we explain how electronic spectra are simulated in the \dirac~package in the framework of time-dependent response theory. In Appendix \ref{sec:Multipolar-gauge} we discuss multipolar gauge and,
contrary to previous works, discuss the gauge transformation between different expansion points. Finally, we present the trivariate beta function which plays a key role in the fully analytic approach to rotational averaging in Appendix \ref{sec:The-trivariate-beta}.

\section{\label{sec:level1}Theory}
We start by reviewing the theory of interactions of molecules with electromagnetic radiation within a relativistic but semi-classical description before deriving three different schemes for computing oscillator strengths beyond the ED approximation. Finally, we present expressions for isotropically averaged oscillator strengths for each case.
The resulting expressions have been implemented in a development version of the \dirac~program.\cite{DIRAC19} 

\subsection{Coupling particles and fields\label{subsec:Coupling-particles-and}}

External fields are introduced into the Hamiltonian $\hat{H}$ through
the substitutions
\begin{equation}
\hat{\mathbf{p}}\rightarrow\hat{\boldsymbol{\pi}}=\hat{\mathbf{p}}-q\mathbf{A};\quad\hat{H}\rightarrow\hat{H}+q\phi,\label{eq:minimal_substitution}
\end{equation}
where appears particle charge $q$, the scalar potential $\phi$,
the vector potential $\mathbf{A}$, linear momentum $\hat{\mathbf{p}}$
and the mechanical momentum $\hat{\boldsymbol{\pi}}$. The expectation
value of the resulting interaction Hamiltonian may then be expressed
as
\begin{equation}
\left\langle \hat{H}_{int}\right\rangle =\int\rho\left(\mathbf{r},t\right)\phi\left(\mathbf{r},t\right)d^{3}\mathbf{r}-\int\mathbf{A}\left(\boldsymbol{r},t\right)\cdot\mathbf{j}\left(\mathbf{r},t\right)d^{3}\mathbf{r},\label{eq:Hint}
\end{equation}
where the scalar potential is seen to couple to the charge density
$\rho$ and the vector potential to the current density $\boldsymbol{j}$.
The substitutions in Eq.~\eqref{eq:minimal_substitution} have been termed the principle of minimal electromagnetic coupling\cite{gell-mann:1956} since it only refers to a single property of the particles, namely charge.
Interestingly, it arises from the interaction Lagrangian proposed
by Schwarzschild\cite{schwarzschild:eldyn} in 1903, two years before the \emph{annus mirabilis}
of Einstein. The expectation value of
the interaction Hamiltonian, Eq.~\eqref{eq:Hint}, can be expressed compactly in terms of 4-current $j_{\mu}$ and 4-potential $A_{\mu}$
\begin{align}
\left\langle \hat{H}_{int}\right\rangle =-\int j_{\mu}\left(\mathbf{r},t\right)A_{\mu}\left(\mathbf{r},t\right)d^{3}\mathbf{r};\quad\begin{cases}
j_{\mu}= & \left(\mathbf{j},ic\rho\right)\\
A_{\mu}= & \left(\mathbf{A},\frac{i}{c}\phi\right)
\end{cases}
\end{align}
($c$ is the speed of light), thus manifestly demonstrating its relativistic
nature. In fact, one may very well argue that in the non-relativistic
limit electrodynamics reduces to electrostatics and that magnetic
induction, in addition to retardation, is a relativistic effect. \cite{saue_AdvQuantChem2005}
Yet, the minimal substitution is
customarily employed also in calculations denoted ``non-relativistic''.
Such calculations in reality use a non-relativistic description of
particles, but a relativistic treatment of their coupling to external
electromagnetic fields. This is perfectly justified from a pragmatic
point of view, but it should be kept in mind that if the sources of
the electromagnetic waves were to be included in the system under
study, their magnetic component would vanish. 

A point we would like to emphasize in the present work is that the
non-relativistic use of the minimal substitution in Eq.~\eqref{eq:minimal_substitution}
leads to a more complicated formalism than the fully relativistic
approach, since the former mixes theories of different transformation
properties. This can be seen by comparing the non-relativistic
and relativistic Hamiltonian operators obtained by minimal substitution. We
may write the non-relativistic free-electron Hamiltonian in two different
forms
\begin{equation}
\hat{h}_{0}^{NR}=\frac{\hat{p}^{2}}{2m_{e}}=\frac{\left(\boldsymbol{\sigma}\cdot\hat{\mathbf{p}}\right)^2}{2m_{e}},\label{eq:NR free-electron}
\end{equation}
where appears the electron mass $m_{e}$ and the Pauli spin matrices
$\boldsymbol{\sigma}$. These two forms are equivalent as long as
external fields are not invoked. Upon minimal substitution, 
one obtains
\begin{equation}
\hat{h}^{NR}=\frac{\hat{p}^{2}}{2m_{e}}+\frac{e}{2m_{e}}\left(\mathbf{p}\cdot\mathbf{A}+\mathbf{A}\cdot\mathbf{p}\right)+\frac{e^{2}A^{2}}{2m_{e}}-e\phi+\frac{e\hbar}{2m_{e}}\left(\boldsymbol{\sigma}\cdot\mathbf{B}\right),\label{eq:NRham}
\end{equation}
where $-e$ is the electron charge and $\hbar$ is the reduced Planck constant.
The final term in Eq.~\eqref{eq:NRham}, representing spin-Zeeman interaction, only appears
if one starts from the second form of Eq.~\eqref{eq:NR free-electron}.
Spin can be thought of as hidden in the non-relativistic free-electron
Hamiltonian. On the other hand, the first form of Eq.~\eqref{eq:NR free-electron} can be thought of as a manifestation of the economy of Nature's laws: neither charge nor spin is required for the description of the free
electron. We may contrast the non-relativistic Hamiltonian (Eq.~\eqref{eq:NRham})
with its relativistic counterpart
\begin{equation}
\hat{h}^{R}=\beta m_{e}c^{2}+c\left(\boldsymbol{\alpha}\cdot\hat{\mathbf{p}}\right)+ec\left(\boldsymbol{\alpha}\cdot\mathbf{A}\right)-e\phi,\label{eq:Rham}
\end{equation}
where appears the Dirac matrices $\boldsymbol{\alpha}$ and $\beta$. Here, the three terms describing magnetic interaction in the non-relativistic framework has
been reduced to a single one, which is linear in the vector potential.
 
\subsection{Full light-matter interaction\label{subsec:Full-light-matter-interaction}}

The Beer--Lambert law, 
\begin{align}
I=I_{0}e^{-{\cal N}\sigma l},
\end{align}
expresses the attenuation of the intensity $I_{0}$ of incoming light
in terms of the \textit{effective} number of absorbing molecules, given as
the product of the number density ${\cal N}$ of absorbing molecules,
the length $l$ of the sample and the absorption cross section $\sigma$.
To find an expression for the absorption cross section, we start from
two equivalent expressions for the rate of energy exchange between
(monochromatic) light and molecules: i) as intensity times
absorption cross section $\sigma$ or ii) as photon energy $\hbar\omega$
times the transition rate $w_{f\leftarrow i}$, that is 
\begin{align}
I(\omega)\sigma\left(\omega\right)=\hbar\omega w_{f\leftarrow i}(\omega).
\end{align}
The intensity is expressed in terms of the electric constant $\varepsilon_{0}$
and the electric field strength $E_{\omega}$
\begin{align}
I(\omega)=\frac{1}{2}\varepsilon_{0}cE_{\omega}^{2}.
\end{align}
Starting from a time-dependent interaction operator of the form
\begin{align}\label{eq:interaction operator}
\hat{V}\left(t\right)=\hat{V}\left(\omega\right)e^{-i\omega t}+\hat{V}\left(-\omega\right)e^{+i\omega t};\quad\hat{V}\left(-\omega\right)=\hat{V}^{\dagger}\left(\omega\right),
\end{align}
an expression for the transition rate $w_{f\leftarrow i}$ may be
found from time-dependent perturbation theory \cite{nrsbook}
\begin{align}
w_{f\leftarrow i}\left(\omega\right)=\frac{2\pi}{\hbar^{2}}\left|\langle f|\hat{V}(\omega_{fi})|i\rangle\right|^{2}f\left(\omega,\omega_{fi},\gamma_{fi}\right).
\end{align}
This formula is often referred to as Fermi's golden rule.
However, the rule actually pertains to transition from a discrete state to continuum of states (see for instance Ref. \citenum{Atkins:MQM1996}), but may be applied to a discrete final state, provided it has a finite lifetime,\cite{cohen-tann2011} here manifested by the lineshape function $f\left(\omega,\omega_{fi},\gamma_{fi}\right)$. Setting 
\begin{align}\label{eq:effT}
\hat{V}\left(\omega\right)=-\frac{1}{2}E_{\omega}\hat{T}\left(\omega\right)
\end{align}
 gives an expression for the absorption cross section
\begin{equation}
\sigma\left(\omega\right)=\frac{\pi\omega}{\varepsilon_{0}\hbar c}\left|\langle f|\hat{T}(\omega_{fi})|i\rangle\right|^{2}f\left(\omega,\omega_{fi},\gamma_{fi}\right),\label{eq:absorption cross section}
\end{equation}
 in terms of an \emph{effective} interaction operator $\hat{T}(\omega)$ (see below).
Closely related is the oscillator strength, defined as
\begin{equation}
f_{fi}\left(\omega\right)=\frac{2\omega}{\hbar e^{2}}\left|\langle f|\hat{T}(\omega_{fi})|i\rangle\right|^{2}f\left(\omega,\omega_{fi},\gamma_{fi}\right).\label{eq:oscillator strength}\end{equation}

In this work, we consider linearly polarized monochromatic light
with electric and magnetic components
\begin{align}
\begin{aligned}\label{eq:linear polarization}
\mathbf{E}\left(\mathbf{r},t\right)=&E_{\omega}\boldsymbol{\epsilon}\sin\left[\mathbf{k}\cdot\mathbf{r}-\omega t+\delta\right]\\
\mathbf{B}\left(\mathbf{r},t\right)=&\frac{E_{\omega}}{\omega}\left(\mathbf{k}\times\boldsymbol{\epsilon}\right)\sin\left[\mathbf{k}\cdot\mathbf{r}-\omega t+\delta\right],
\end{aligned}
\end{align}
where appears the wave vector $\mathbf{k}$ with length
\begin{equation}\label{eq:k}
  k=\frac{\omega}{c}=\frac{2\pi}{\lambda},
  \end{equation}
the polarization vector
$\boldsymbol{\epsilon}$ and the phase $\delta$. Such an electromagnetic
wave is conventionally represented in Coulomb (radiation) gauge by the scalar and vector potentials
\begin{equation}
\tilde{\phi}\left(\mathbf{r},t\right)=0;\quad\tilde{\mathbf{A}}\left(\mathbf{r},t\right)=-\frac{E_{\omega}}{\omega}\boldsymbol{\epsilon}\cos\left[\mathbf{k}\cdot\mathbf{r}-\omega t+\delta\right].\label{eq:linear plane wave potentials}
\end{equation}
 Starting from the Dirac Hamiltonian in Eq.~\eqref{eq:Rham}, this
leads to an effective interaction operator of the form
\begin{equation}\label{eq:reduced interaction operator}
\hat{T}_{\mathrm{full}}\left(\omega\right)=\frac{e}{\omega}\left(c\boldsymbol{\alpha}\cdot\boldsymbol{\epsilon}\right)e^{+i\left(\mathbf{k}\cdot\mathbf{r}+\delta\right)};\quad
\hat{T}_{\mathrm{full}}^{\dagger}\left(\omega\right)=\hat{T}_{\mathrm{full}}\left(-\omega\right).
\end{equation}
It is clear from Eq.~\eqref{eq:NRham} that the corresponding effective
interaction operator in the non-relativistic framework will have a
more complicated expression. However, simplifications are introduced
by invoking a weak-field approximation such that the third term, the
diamagnetic contribution, is neglected. Also, the fourth term, the
spin-Zeeman contribution, is often ignored.

One straightforwardly establishes that use of the full interaction operator assures gauge-origin independence of intensities.\cite{List_JCP2015} Upon a change of gauge-origin $\mathbf{O}\rightarrow\mathbf{O}+\mathbf{a}$ a constant complex phase is introduced in the interaction operator
\begin{equation}\label{eq:Tgauge}
\hat{T}_{\mathrm{full}}\left(\omega;\mathbf{O}\right)\rightarrow\hat{T}_{\mathrm{full}}\left(\omega;\mathbf{O}+\mathbf{a}\right)=\hat{T}_{\mathrm{full}}\left(\omega;\mathbf{O}\right)e^{+i\left(\mathbf{k}\cdot\mathbf{a}\right)}.
\end{equation}
This phase is, however, cancelled by its complex conjugated partner when the interaction operator is inserted into the expressions for the absorption cross section, Eq.~\eqref{eq:absorption cross section}, or oscillator strength, Eq.~\eqref{eq:oscillator strength}.

The ED approximation assumes that the dimensionless quantity $\left\langle kr\right\rangle \ll1$
such that the interaction operator may be approximated by
\begin{align}
\hat{T}_{V}\left(\omega\right)=\frac{e}{\omega}\left(c\boldsymbol{\alpha}\cdot\boldsymbol{\epsilon}\right)e^{+i\delta}\quad\xrightarrow{\delta=0}\quad\frac{e}{\omega}\left(c\boldsymbol{\alpha}\cdot\boldsymbol{\epsilon}\right),\label{eq: velocity representation}
\end{align}
which physically corresponds to the absorbing molecule effectively
seeing a uniform electric field. The subscript $'V'$ refers to the
velocity representation. To convert to the length representation we
use the following expression for the velocity operator
\begin{align}
\hat{\mathbf{v}}=-\frac{i}{\hbar}\left[\mathbf{r},\hat{h}\right],
\end{align}
obtained from the Heisenberg equation of motion, an observation that can be traced back at least to the first edition (1930) of Dirac's monograph.\cite{Dirac:1930:PQM} In the non-relativistic
case, this leads to a velocity operator of the form $\hat{\mathbf{v}}^{\mathrm{NR}}=\hat{\mathbf{p}}/m_{e}$,
which is straightforwardly related to the corresponding classical expression.
In the relativistic case, one obtains the less intuitive form\cite{Breit_PNAS1928,saue_CPC2011}  $\mathbf{v}^{\mathrm{R}}=c\boldsymbol{\alpha}$, expressing the \textit{Zitterbewegung} of the electron, 
which facilitates the connection
\begin{align}
&\langle f|\hat{T}_{V}\left(\omega\right)|i\rangle=\langle f|\hat{T}_{L}\left(\omega\right)|i\rangle\label{eq:ED_v2l}\\[0.1in]
&\hat{T}_{L}\left(\omega\right)=-ie^{+i\delta}\left(\frac{\omega_{fi}}{\omega}\right)\hat{\boldsymbol{\mu}}\cdot\boldsymbol{\epsilon}\quad\xrightarrow[\omega=\omega_{fi}]{\delta=\pi/2}\quad\hat{\boldsymbol{\mu}}\cdot\boldsymbol{\epsilon};\quad\hat{\boldsymbol{\mu}}=-e\mathbf{r}.\label{eq:length representation}
\end{align}
We prefer to refer to these forms as \textit{representations} rather than gauges (see also Ref.\citenum{Bauschlicher_TCA1991} and references therein). Gauge freedom arises from the observation that the longitudinal component of the vector potential does not contribute to the magnetic field, and gauges are accordingly fixed by imposing conditions on this component. For instance, the condition $\boldsymbol{\nabla}\cdot\mathbf{A}=0$ of Coulomb gauge states that the longitudinal component of the vector potential is zero. Although the underlying potentials of the length and velocity representations are related by a gauge transformation, there is, as far as we can see, no gauge condition separating them. Both satisfy Coulomb gauge, but this is no longer the case for the length representation when going beyond the ED approximation, as demonstrated in Appendix \ref{sec:Multipolar-gauge}.

At this point it should be noted that whereas the time-dependent effective
interaction operator $\hat{T}_{\mathrm{full}}\left(t\right)$ is necessarily
Hermitian, this is generally not the case for the frequency-dependent
component $\hat{T}_{\mathrm{full}}\left(\omega\right)$, as seen from Eq.~\eqref{eq:reduced interaction operator}. We shall, however,
insist that the effective interaction operators are Hermitian within
the ED approximation. This leads to the following choices
for the phase $\delta$ of the electromagnetic plane wave, Eq.~\eqref{eq:linear polarization}, 
\begin{equation}
\delta=\begin{cases}
0; & \mbox{(velocity representation)}\\
\pi/2; & \mbox{(length representation)}
\end{cases}.\label{eq:phase convention}
\end{equation}

In the present work, we report the implementation of three different schemes for simulation of electronic spectra beyond the ED approximation within a linear response framework.
 More details about the underlying theory and the implementation are given in Appendix \ref{sec:Simulation-of-electronic}.
Two features of the present implementation of the full light-matter interaction operator should be stressed: i) integrals
over the effective interaction operator, Eq.~\eqref{eq:reduced interaction operator}, 
in a Gaussian basis are identified as Fourier transforms with simple
analytic expressions,\cite{List_JCP2015} and ii) the effective interaction
operator, Eq.~\eqref{eq:reduced interaction operator}, is a general operator, and thus it may be split into Hermitian and anti-Hermitian parts
\begin{align}
\hat{T}_{H}\left(\omega\right)=&\frac{e}{\omega}\left(c\boldsymbol{\alpha}\cdot\boldsymbol{\epsilon}\right)\cos\left(\mathbf{k}\cdot\mathbf{r}\right)\\
\hat{T}_{A}\left(\omega\right)=&\frac{e}{\omega}\left(ic\boldsymbol{\alpha}\cdot\boldsymbol{\epsilon}\right)\sin\left(\mathbf{k}\cdot\mathbf{r}\right).
\end{align}
 The Hermitian and anti-Hermitian operators are time-antisymmetric
and time-symmetric, respectively. In accordance with the quaternion
symmetry scheme of \dirac\cite{saue_JCP1999} an imaginary $i$ will be inserted in the
Hermitian part to make it time-symmetric. The components can be further
broken down on spatial symmetries using
\begin{align}
\begin{array}{cccrl}
e^{\pm i\left(\mathbf{k}\cdot\mathbf{r}\right)} & = & \cos\left(k_{x}x\right)\cos\left(k_{y}y\right)\cos\left(k_{z}z\right) &  & \left(\Gamma_{0}\right)\\
 & - & \sin\left(k_{x}x\right)\sin\left(k_{y}y\right)\cos\left(k_{z}z\right) &  & \left(\Gamma_{R_{z}}\right)\\
 & - & \sin\left(k_{x}x\right)\cos\left(k_{y}y\right)\sin\left(k_{z}z\right) &  & \left(\Gamma_{R_{y}}\right)\\
 & - & \cos\left(k_{x}x\right)\sin\left(k_{y}y\right)\sin\left(k_{z}z\right) &  & \left(\Gamma_{R_{x}}\right)\\
 & \mp i & \sin\left(k_{x}x\right)\sin\left(k_{y}y\right)\sin\left(k_{z}z\right) &  & \left(\Gamma_{xyz}\right)\\
 & \pm i & \cos\left(k_{x}x\right)\cos\left(k_{y}y\right)\sin\left(k_{z}z\right) &  & \left(\Gamma_{z}\right)\\
 & \pm i & \cos\left(k_{x}x\right)\sin\left(k_{y}y\right)\cos\left(k_{z}z\right) &  & \left(\Gamma_{y}\right)\\
 & \pm i & \sin\left(k_{x}x\right)\cos\left(k_{y}y\right)\cos\left(k_{z}z\right) &  & \left(\Gamma_{x}\right).
\end{array}
\end{align}
Here $\Gamma_{0}$ refers to the totally symmetric irrep, $\left(\Gamma_{x,}\Gamma_{y},\Gamma_{z}\right)$
to the symmetries of the coordinates, $\left(\Gamma_{R_{x}},\Gamma_{R_{y}},\Gamma_{R_{z}}\right)$
to the symmetry of the rotations and $\Gamma_{xyz}$ to the symmetry
of the function $xyz$. Together, these eight symmetries form the
eight irreps of the $D_{2h}$ point group, whereas some symmetries
coalesce for subgroups. In the present implementation, for an excitation of given (boson) symmetry, we only invoke the relevant contribution from $e^{\pm i\left(\mathbf{k}\cdot\mathbf{r}\right)}$.

\subsection{Truncated light-matter interaction\label{subsec:truncated_light_matter}}

In this section, we derive expressions for the absorption cross section
or oscillator strength truncated to finite order in the length of
the wave vector. In the first subsection, we develop a compact formalism based
directly on an expansion of the effective interaction operator, Eq.~\eqref{eq:reduced interaction operator}.
Next, we provide the relativistic extension of the theory developed by Bernadotte and co-workers,\cite{bernadotte_JCP2012}
where oscillator strengths are expressed in terms of electric and magnetic multipoles.
We shall, however, obtain these expressions in a more straightforward manner by using multipolar gauge. 
The two approaches can to some extent be thought of as generalizations
of the velocity  and length representation, respectively, to arbitrary
orders in the wave vector.

\subsubsection{Coulomb gauge: velocity representation\label{subsec:Coulomb gauge}}

A direct approach for obtaining the absorption cross section
(or oscillator strength) to some order in the wave vector is to perform
a Taylor-expansion of the absorption cross section in Eq.~\eqref{eq:absorption cross section}
in orders of the wave vector, that is
\begin{align}\label{eq:gen_vel_gauge}
\sigma\left(\omega\right)  &=  \frac{\pi\omega}{\varepsilon_{0}\hbar c}\sum_{n=0}^{\infty}\frac{k^{n}}{n!}\frac{d^{n}}{dk^{n}}\left[\langle f|\hat{T}_{\mathrm{full}}(\omega_{fi})|i\rangle\langle f|\hat{T}_{\mathrm{full}}(\omega_{fi})|i\rangle^{\ast}\right]_{k=0}f\left(\omega,\omega_{fi},\gamma_{fi}\right)\notag\\
 & =  \frac{\pi\omega}{\varepsilon_{0}\hbar c}\sum_{n=0}^{\infty}\sum_{m=0}^{n}\langle f|\hat{T}_{\mathrm{full}}^{\left[n-m\right]}(\omega_{fi})|i\rangle\langle f|\hat{T}_{\mathrm{full}}^{\left[m\right]}(\omega_{fi})|i\rangle^{\ast}f\left(\omega,\omega_{fi},\gamma_{fi}\right),
\end{align}
where appears Taylor coefficients 
\begin{align}\label{eq:T_n}
\hat{T}_{\mathrm{full}}^{\left[n\right]}(\omega)=\frac{k^{n}}{n!}\frac{d^{n}}{dk^{n}}\left[\frac{e}{\omega}\left(c\boldsymbol{\alpha}\cdot\boldsymbol{\epsilon}\right)e^{+i\left(\mathbf{k}\cdot\mathbf{r}\right)}\right]_{k=0}=\frac{e}{\omega}\frac{1}{n!}\left(c\boldsymbol{\alpha}\cdot\boldsymbol{\epsilon}\right)\left(i\mathbf{k}\cdot\mathbf{r}\right)^{n}
\end{align}
in the corresponding expansion of the effective interaction operator, 
Eq.~\eqref{eq:reduced interaction operator}, with the phase $\delta=0$,
according to the phase convention in Eq.~\eqref{eq:phase convention}.
From inspection we find that even- and odd-order operators are time-antisymmetric and time-symmetric, respectively. It should be noted that the underlying, truncated vector potential satisfies Coulomb gauge.

We may separate the absorption cross section into even- and odd-order contributions with respect to the wave vector, that is
\begin{align}
\sigma^{\left[2n\right]}\left(\omega\right)&=\frac{\pi\omega}{\varepsilon_{0}\hbar c}\sum_{m=0}^{n}\left(2-\delta_{m0}\right)\mathrm{Re}\left\{ \langle f|\hat{T}_{\mathrm{full}}^{\left[n-m\right]}(\omega_{fi})|i\rangle\langle f|\hat{T}_{\mathrm{full}}^{\left[n+m\right]}(\omega_{fi})|i\rangle^{\ast}\right\} f\left(\omega,\omega_{fi},\gamma_{fi}\right) \label{eq:osc_vr}\\
\sigma^{[2n+1]}\left(\omega\right) & = \frac{\pi\omega}{\varepsilon_{0}\hbar c}\sum_{m=0}^{n}2\mathrm{Re}\left\{ \langle f|\hat{T}_{\mathrm{full}}^{\left[n+m+1\right]}(\omega_{fi})|i\rangle\langle f|\hat{T}_{\mathrm{full}}^{\left[n-m\right]}(\omega_{fi})|i\rangle^{\ast}\right\} f\left(\omega,\omega_{fi},\gamma_{fi}\right)=0. \label{eq:osc_vr_odd}
\end{align}
The odd-order contributions vanish identically because the two interaction operators of each term,
contrary to the even-order terms, will have opposite symmetry with respect to time reversal, such that the product of their transition moments will be imaginary (see Appendix \ref{sec:Simulation-of-electronic}).

The demonstration of formal gauge-origin independence in the generalized velocity representation at each order $n$ in the wavevector follows straightforwardly from Eq.~\eqref{eq:gen_vel_gauge}, being $n$th order derivatives of a term that is gauge-origin independent for all values of $k$. This result was obtained earlier in the non-relativistic framework by Bernadotte \textit{et al.}~but in a somewhat elaborate manner (see Appendix C of Ref.~\citenum{bernadotte_JCP2012}). Their derivation, however, highlights the challenge of achieving gauge-origin independence in practical calculations, and so we shall give a slightly more compact version here: Upon a change of gauge-origin $\mathbf{O}\rightarrow\mathbf{O}+\mathbf{a}$, the $n$th order interaction operator
in the velocity representation may be expressed as
\begin{equation}\label{eq:Tn_gauge}
\hat{T}^{[n]}_{\mathrm{full}}(\omega;\mathbf{O})\rightarrow\hat{T}_{\mathrm{full}}^{[n]}(\omega;\mathbf{O}+\mathbf{a})=\sum_{m=0}^{n}\frac{1}{m!}\left(i\mathbf{k}\cdot\mathbf{a}\right)^{m}\hat{T}_{\mathrm{full}}^{[n-m]}(\omega;\mathbf{O}),
\end{equation}
which follows from Eq.~\eqref{eq:Tgauge}. The $n$th order absorption cross section at the new gauge origin may then be expressed as
\begin{eqnarray}
\sigma^{\left[n\right]}\left(\omega;\mathbf{O}+\mathbf{a}\right) & = & \sum_{m=0}^{n}\langle f|T_{\mathrm{full}}^{[n-m]}\left(\omega;\mathbf{O}+\mathbf{a}\right)|i\rangle\langle f|T_{\mathrm{full}}^{[m]}\left(\omega;\mathbf{O}+\mathbf{a}\right)|i\rangle^{\ast}\label{eq:Sgauge}\\
 & = & \sum_{m=0}^{n}\sum_{p=0}^{n-m}\sum_{q=0}^{m}\frac{1}{p!q!}\left(-1\right)^{q}\left(i\mathbf{k}\cdot\mathbf{a}\right)^{p+q}\langle f|T_{\mathrm{full}}^{[n-m-p]}\left(\omega;\mathbf{O}\right)|i\rangle\langle f|T_{\mathrm{full}}^{[m-q]}\left(\omega;\mathbf{O}\right)|i\rangle^{\ast}\nonumber
\end{eqnarray}
If we take the orders for each pair of interaction operators as indices of a matrix, we see that the
pairs $(n-m,m)$ from the first line fills the antidiagonal of a square matrix with indices running from 0 to $n$, whereas the pairs
$(n-m-p,n-q)$ from the second line fills the triangle above as well. This suggests to replace indices $m$ and $p$ by $u=n-m-p$ and $v=m-q$. After rearrangement, this leads to the expression
\begin{equation}
\sigma^{\left[n\right]}\left(\omega;\mathbf{O}+\mathbf{a}\right)=\sum_{v=0}^{n}\sum_{u=0}^{n-v}\langle f|T_{\mathrm{full}}^{[u]}\left(\omega;\mathbf{O}\right)|i\rangle\langle f|T_{\mathrm{full}}^{[v]}\left(\omega;\mathbf{O}\right)|i\rangle^{\ast}\frac{\left(i\mathbf{k}\cdot\mathbf{a}\right)^{n-(u+v)}}{(n-\left(u+v\right))!}\times{\cal M},
\end{equation}
where
\begin{equation}
  {\cal M}=\sum_{q=0}^{n-(u+v)}\left(\begin{array}{c}
(n-\left(u+v\right))\\q\end{array}\right)\left(-1\right)^{q}=(1-1)^{n-(u+v)}.
\end{equation}
The factor ${\cal M}$ is zero unless $u=n-v$ which reduces the second line of Eq.~\eqref{eq:Sgauge} to the same form as the first line, hence demonstrating that $\sigma^{\left[n\right]}\left(\omega;\mathbf{O}+\mathbf{a}\right)=\sigma^{\left[n\right]}\left(\omega;\mathbf{O}\right)$, that is, the oscillator strengths are indeed gauge-origin independent.
However, one should note that the lower-order interaction operators introduced in Eq.~\eqref{eq:Tn_gauge} upon a change of gauge origin involve multiplication with powers of the displacement as well as the wave vector. This may eventually introduce numerical issues, as will be shown in Section \ref{sec:origindep}.

\subsubsection{Multipolar gauge: length representation}\label{subsec:mg}
A very convenient way of introducing electric and magnetic multipoles
is through the use of multipolar gauge,\cite{Kobe_AJP1980,Stewart_JPA1999,saue2002relprop,nrsbook}
also known as Bloch gauge,\cite{bloch:gauge,Lazzeretti_TCA1993} Barron--Gray gauge\cite{Barron_Gray_JPA1973} or 
Poincar\'e gauge,\cite{Brittin_AJP1982,Skagerstam_AJP1983,cohen:qed,Jackson_RMP2001}
reflecting a history of multiple rediscoveries. In Appendix \ref{sec:Multipolar-gauge},
we provide a compact derivation of the multipolar gauge, avoiding
excessive use of indices. In multipolar gauge the potentials are given
in terms of the electric and magnetic fields and their derivatives
at some expansion point $\mathbf{a}$. 
When inserted into the interaction
Hamiltonian in Eq.~\eqref{eq:Hint}, they automatically provide an expansion
of the light-matter interaction in terms of electric and magnetic
multipoles of the molecule. 

We first consider the form of the effective interaction in multipolar gauge,
starting from the electromagnetic plane wave, Eq.~\eqref{eq:linear polarization}, 
represented by the potentials in Eq.~\eqref{eq:linear plane wave potentials}.
In accordance with the discussion in Section \ref{subsec:Full-light-matter-interaction}
and the phase convention of Eq.~\eqref{eq:phase convention}, we set the
phase of the plane wave to $\delta=\pi/2$. The potentials in multipolar
gauge (mg) are then given by
\begin{align}
\phi\left(\mathbf{r},t\right) & =  -\frac{1}{2}\left(\boldsymbol{\delta}\cdot\mathbf{E}_{\omega}\right)\sum\limits_{n=0}^{\infty}\frac{1}{\left(n+1\right)!}\left\{\left(i\mathbf{k}\cdot\boldsymbol{\delta}\right)^{n}e^{i\left(\mathbf{k}\cdot\mathbf{a}-\omega t\right)}+\left(-i\mathbf{k}\cdot\boldsymbol{\delta}\right)^{n}e^{-i\left(\mathbf{k}\cdot\mathbf{a}-\omega t\right)}\right\};\quad\boldsymbol{\delta}=\mathbf{r}-\mathbf{a}\\
\mathbf{A}\left(\mathbf{r},t\right) & =  -\frac{1}{2}\left(\boldsymbol{\delta}\times\mathbf{B}_{\omega}\right)\sum\limits_{n=1}^{\infty}\frac{n}{\left(n+1\right)!}\left\{\left(i\mathbf{k}\cdot\boldsymbol{\delta}\right)^{n-1}e^{i\left(\mathbf{k}\cdot\mathbf{a}-\omega t\right)}+\left(-i\mathbf{k}\cdot\boldsymbol{\delta}\right)^{n-1}e^{-i\left(\mathbf{k}\cdot \mathbf{a}-\omega t\right)}\right\}.
\end{align}

Setting the expansion point $\mathbf{a}=\mathbf{0}$, we find that the effective interaction operator may be expressed as
\begin{equation}\label{eq:T_mp}
  \hat{T}_{\mathrm{mg}}=\sum_{n=0}^{\infty} \hat{T}^{[n]}_{\mathrm{mg}},
\end{equation}
where
\begin{align}
\hat{T}^{[0]}_{\mathrm{mg}}\left(\omega\right)&=-e\left(\mathbf{r}\cdot\boldsymbol{\epsilon}\right)\\
\hat{T}^{[n]}_{\mathrm{mg}}\left(\omega\right)&=-e\left[\frac{1}{\left(n+1\right)!}\left(\mathbf{r}\cdot\boldsymbol{\epsilon}\right)\left(i\mathbf{k}\cdot\mathbf{r}\right)^{n}-\frac{i}{\omega}\frac{n}{\left(n+1\right)!}\left(i\mathbf{k}\times\boldsymbol{\epsilon}\right)\cdot\left(\mathbf{r}\times c\boldsymbol{\alpha}\right)\left(i\mathbf{k}\cdot\mathbf{r}\right)^{n-1}\right],\quad n\ne 0.\label{eq:Tmulti}
\end{align}
Further insight is obtained by writing the effective interaction operator on component form as
\begin{equation}
\hat{T}_{\mathrm{mg}}\left(\omega\right)=\hat{Q}_{p}^{\left[1\right]}\epsilon_{p}+\sum_{n=1}^{\infty}i^{n}\epsilon_{p}k_{j_{1}}k_{j_{2}}\ldots k_{j_{n}}\hat{X}_{j_{1}\ldots j_{n};p}^{[n]}\left(\omega\right).\label{eq:mpol interaction}
\end{equation}
In the above expression, we employ the Einstein summation convention
and introduce the multipole operator $\hat{X}^{[n]}$ associated with
$O\left(k^{n}\right)$ 
\begin{equation}
\hat{X}_{j_{1}\ldots j_{n};p}^{[n]}\left(\omega\right)=\frac{1}{\left(n+1\right)!}\hat{Q}_{j_{1}\ldots j_{n},p}^{\left[n+1\right]}-\frac{i}{\omega}\frac{1}{n!}\hat{m}_{j_{1}\ldots j_{n-1};r}^{\left[n\right]}\varepsilon_{rj_{n}p},\label{eq:multipole operator}
\end{equation}
where appears the Levi--Civita symbol $\varepsilon_{ijk}$. This operator
is in turn built from the electric and magnetic multipole operators
\begin{align}
\hat{Q}_{j_{1}\ldots j_{n}}^{\left[n\right]}&=-er_{j_1}r_{j_2}\ldots r_{j_{n}}\\
\quad\hat{m}_{j_{1}\ldots j_{n-1};j_{n}}^{\left[n\right]}&=\frac{n}{n+1}r_{j_{1}}r_{j_{2}}\ldots r_{j_{n-1}}(\mathbf{r}\times\hat{\mathbf{j}})_{j_{n}};\quad\hat{\mathbf{j}}=-ec\boldsymbol{\alpha}.
\end{align}
 Again we would like to stress the simplicity of the relativistic
formalism compared to the non-relativistic one: the magnetic multipole
operators $\hat{m}^{\left[n\right]}$ contain the current density
operator $\hat{\mathbf{j}}$ which in the relativistic form is simply
electron charge times the velocity operator, allowing straightforward
implementation of the magnetic multipole operator to arbitrary order. The non-relativistic form is
more involved containing contributions from the mechanical momentum operator as well as the curl of the spin magnetization.\cite{Bast_CP2009}
One may note that the electric and magnetic multipole operators are
time-symmetric and time-antisymmetric, respectively. However, in Eq.~\eqref{eq:multipole operator}
the magnetic multipole operator is multiplied with imaginary $i$
such that the multipole operator $\hat{X}^{[n]}$ is time-symmetric,
fitting well into the quaternion symmetry scheme of \dirac{}. 

Inserting the effective interaction operator $\hat{T}_{\mathrm{mg}}\left(\omega\right)$
into the expression for the absorption cross section in Eq.~\eqref{eq:absorption cross section}
and expanding in orders of the wave vector, we find that odd-order contributions
to the absorption cross section vanish, as was also the case in the velocity representation, whereas the even-order ones may be expressed as
\begin{align}
\sigma^{[2n]}\left(\omega\right)  &= \frac{\pi\omega}{\varepsilon_{0}\hbar c}\sum_{m=0}^{n}\left(-1\right)^{m}\left(2-\delta_{m0}\right)\epsilon_{p}\epsilon_{q}k_{j_{1}}k_{j_{2}}\ldots k_{j_{2n}}\notag\\
&\times\text{Re}\left\{ \langle f|\hat{X}_{j_{1}\ldots j_{n+m};p}^{[n+m]}\left(\omega\right)|i\rangle\langle f|\hat{X}_{j_{n+m+1}\ldots j_{2n};q}^{[n-m]}\left(\omega\right)|i\rangle^{\ast}\right\} f\left(\omega,\omega_{fi},\gamma_{fi}\right).
\end{align}

We may connect the interaction operators of multipolar gauge with those of Coulomb gauge in the velocity representation. Starting from Eq.~\eqref{eq:T_n}, we use the relation
\begin{equation}\label{eq:useful}
\left(i\mathbf{k}\times\boldsymbol{\epsilon}\right)\cdot\left(\mathbf{r}\times c\boldsymbol{\alpha}\right)=\left(c\boldsymbol{\alpha}\cdot\boldsymbol{\epsilon}\right)\left(i\mathbf{k}\cdot\mathbf{r}\right)-\left(\mathbf{r}\cdot\boldsymbol{\epsilon}\right)\left(i\mathbf{k}\cdot c\boldsymbol{\alpha}\right)
\end{equation}
to obtain
\begin{align}
  T^{[n]}\left(\omega\right)
  &=\frac{e}{\omega}\frac{1}{\left(n+1\right)!}\left\{ \left(c\boldsymbol{\alpha}\cdot\boldsymbol{\epsilon}\right)\left(i\mathbf{k}\cdot\mathbf{r}\right)^{n}+n\left(\mathbf{r}\cdot\boldsymbol{\epsilon}\right)\left(i\mathbf{k}\cdot c\boldsymbol{\alpha}\right)\left(i\mathbf{k}\cdot\mathbf{r}\right)^{n-1}\right\}\nonumber\\ &+\frac{e}{\omega}\frac{n}{\left(n+1\right)!} \left(i\mathbf{k}\times\boldsymbol{\epsilon}\right)\cdot\left(\mathbf{r}\times c\boldsymbol{\alpha}\right)\left(i\mathbf{k}\cdot\mathbf{r}\right)^{n-1}.
\end{align}  
Comparing with Eq.~\eqref{eq:Tmulti}, we see that the second term above contains the $n$th-order magnetic multipole operator, which implies that the we may extract from the first term the $(n+1)$th-order electric multipole operator in the \textit{velocity representation}.
Next we use
\begin{equation}\label{eq:Tcomm}
-\frac{i}{\hbar}\left[\left(\mathbf{r}\cdot\boldsymbol{\epsilon}\right)\left(i\mathbf{k}\cdot\mathbf{r}\right)^{n},\hat{h}\right]=\left(c\boldsymbol{\alpha}\cdot\boldsymbol{\epsilon}\right)\left(i\mathbf{k}\cdot\mathbf{r}\right)^{n}+n\left(\mathbf{r}\cdot\boldsymbol{\epsilon}\right)\left(i\mathbf{k}\cdot c\boldsymbol{\alpha}\right)\left(i\mathbf{k}\cdot\mathbf{r}\right)^{n-1}
\end{equation}
to arrive at
\begin{equation}
T^{[n]}\left(\omega\right)=\frac{-ie}{\hbar\omega}\frac{1}{\left(n+1\right)!}\left[\left(\mathbf{r}\cdot\boldsymbol{\epsilon}\right)\left(i\mathbf{k}\cdot\mathbf{r}\right)^{n},\hat{h}\right]+\frac{e}{\omega}\frac{n}{\left(n+1\right)!}\left(i\mathbf{k}\times\boldsymbol{\epsilon}\right)\cdot\left(\mathbf{r}\times c\boldsymbol{\alpha}\right)\left(i\mathbf{k}\cdot\mathbf{r}\right)^{n-1} 
\end{equation}
In order to complete the derivation, we have to form transition moments, which provide the connection
\begin{equation}\label{eq:BED_v2l}
  \langle f|\hat{T}_{\mathrm{full}}^{\left[n\right]}(\omega)|i\rangle = -i\langle f|\hat{T}_{\mathrm{mg}}^{\left[n\right]}(\omega)|i\rangle
\end{equation}
between velocity and length representations and generalizes Eq.~\eqref{eq:ED_v2l} to arbitrary order in the wave vector; in Eq.~\eqref{eq:ED_v2l} the negative imaginary phase is cancelled by choosing the phase $\delta=\pi/2$, which is not done here. At this point one should note that the derivation is greatly simplified by the fact that the relativistic velocity operator $c\boldsymbol{\alpha}$ commutes with the coordinates, contrary to the non-relativistic one. Furthermore, as discussed at the end of Appendix \ref{sec:Multipolar-gauge}, the appearance of a commutator involving the Hamiltonian can be taken as an indication of a gauge transformation and, indeed,
we show that the operator appearing together with the Hamiltonian in Eq.~\eqref{eq:Tcomm} is the
gauge function of multipolar gauge, Eq.~\eqref{eq:mg gauge function}, obtained by inserting the vector potential, Eq.~\eqref{eq:linear plane wave potentials}, of a linear plane wave and retaining the term of order $n$ in the wave vector.

Multipolar gauge has mostly been discussed in the framework of atomic physics where the nuclear origin
provides a natural expansion point. In a molecule there is generally no natural expansion point, and gauge-origin independence becomes an issue.
Starting from Eq.~\eqref{eq:Tn_gauge} and using the connection Eq.~\eqref{eq:BED_v2l}, one straightforwardly derives 
\begin{equation}\label{eq:Tn_mg_gauge}
\langle f|\hat{T}^{[n]}_{\mathrm{mg}}(\omega;\mathbf{O})|i\rangle\rightarrow\langle f|\hat{T}_{\mathrm{mg}}^{[n]}(\omega;\mathbf{O}+\mathbf{a})|i\rangle =\sum_{m=0}^{n}\frac{1}{m!}\left(i\mathbf{k}\cdot\mathbf{a}\right)^{m}\langle f|\hat{T}_{\mathrm{mg}}^{[n-m]}(\omega;\mathbf{O})|i\rangle,
\end{equation}
from which gauge-origin independence of absorption cross sections to all orders in the wave vector follows, using the same demonstration as for Coulomb gauge (velocity representation) in the previous section.
However, the demonstration this time hinges on the connection Eq.~\eqref{eq:BED_v2l}, which is established using commutator relations involving the Hamiltonian that do not necessarily hold in a finite basis
and which effectively amount to a gauge transformation. We have not been able to show gauge-origin independence of absorption cross sections while staying within multipolar gauge, except for the zeroth order term (electric-dipole approximation) where it follows from orthogonality of states. In fact, 
in Appendix \ref{sec:Multipolar-gauge} we show that potentials derived with respect to two different expansion points are related by a gauge transformation, but apparently \textit{only} to the extent that the expansion is not truncated. This suggests that the lack of origin invariance of oscillator strengths observed by Bernadotte \textit{et al.}\cite{bernadotte_JCP2012} and others, including us (see below),
in calculations using an effective interaction operator on multipolar form is more than a finite basis set effect. It also makes sense since truncating the Taylor expansion of electric and magnetic fields inevitably conserves only local information.

\subsection{Rotational averages\label{subsec:rotational_avg}}

\subsubsection{General}

An often encountered experimental situation involves freely rotating
molecules, and we will therefore have to consider rotational averaging.
However, rather than rotating the molecules we shall rotate the experimental
configuration. To this end, we use the unit vectors of the spherical
coordinates
\begin{align}
\begin{aligned}\label{eq:radial unit vector}
\mathbf{e}_{r} & =  \boldsymbol{\mathbf{e}}_{x}\sin\theta\cos\phi+\boldsymbol{\mathbf{e}}_{y}\sin\theta\sin\phi+\boldsymbol{\mathbf{e}}_{z}\cos\theta\\
\boldsymbol{\mathbf{e}}_{\theta} & =  \boldsymbol{\mathbf{e}}_{x}\cos\theta\cos\phi+\boldsymbol{\mathbf{e}}_{y}\cos\theta\sin\phi-\boldsymbol{\mathbf{e}}_{z}\sin\theta \\
\boldsymbol{\mathbf{e}}_{\phi} & =  -\boldsymbol{\mathbf{e}}_{x}\sin\phi+\boldsymbol{\mathbf{e}}_{y}\cos\phi,
\end{aligned}
\end{align}
that reduce to $(\mathbf{e}_{z},\mathbf{e}_{x},\mathbf{e}_{y})$ when
the angles $\theta$ and $\phi$ are both set to zero. More precisely,
we shall align the wave unit vector $\mathbf{e}_{k}$ with the radial
unit vector $\mathbf{e}_{r}$. The polarization vector $\boldsymbol{\epsilon}$
is then in the plane spanned by the unit vectors $\mathbf{e}_{\theta}$
and $\mathbf{e}_{\phi}$. Accordingly we set
\begin{align}
\mathbf{e}_{k}=\mathbf{e}_{r};\quad\boldsymbol{\epsilon}=\cos\chi\boldsymbol{\mathbf{e}}_{\theta}+\sin\chi\boldsymbol{\mathbf{e}}_{\phi},
\end{align}
introducing a third angle $\chi$. The rotational average is defined
as
\begin{align}
\left\langle g\left(\mathbf{r}\right)\right\rangle _{\theta,\phi,\chi}=\frac{1}{8\pi^{2}}\int_{0}^{2\pi}\int_{0}^{2\pi}\int_{0}^{\pi}g\left(\mathbf{r}\right)\sin\theta d\theta d\phi d\chi.
\end{align}

\subsubsection{Full light-matter interaction}

Starting from Eq.~\eqref{eq:absorption cross section} the rotationally
average absorption cross section reads (for any choice of the phase
$\delta$)
\begin{align}
\left\langle \sigma\left(\omega\right)\right\rangle _{\theta,\phi,\chi}=\frac{\pi\omega}{\varepsilon_{0}\hbar c}\left(\frac{e}{\omega_{fi}}\right)^{2}\left\langle \epsilon_{\mu}\epsilon_{\nu}\langle f|c\alpha_{\mu}e^{+i\left(\mathbf{k}\cdot\mathbf{r}\right)}|i\rangle\langle f|c\alpha_{\nu}e^{+i\left(\mathbf{k}\cdot\mathbf{r}\right)}|i\rangle^{\ast}\right\rangle _{\theta,\phi,\chi}f\left(\omega,\omega_{fi},\gamma_{fi}\right).
\end{align}
We first note that the $\chi$-dependence only enters the polarization
vector $\boldsymbol{\epsilon}$, so that we may write
\begin{align}\label{eq:avg BED osc}
\left\langle \sigma\left(\omega\right)\right\rangle _{\theta,\phi,\chi}=\frac{\pi\omega}{\varepsilon_{0}\hbar c}\left(\frac{e}{\omega_{fi}}\right)^{2}\left\langle \left\langle \epsilon_{p}\epsilon_{q}\right\rangle _{\chi}\langle f|c\alpha_{p}e^{+i\left(\mathbf{k}\cdot\mathbf{r}\right)}|i\rangle\langle f|c\alpha_{q}e^{+i\left(\mathbf{k}\cdot\mathbf{r}\right)}|i\rangle^{\ast}\right\rangle _{\theta,\phi}f\left(\omega,\omega_{fi},\gamma_{fi}\right).
\end{align}
 The $\chi$-average has a simple analytic expression in terms of
the components of the radial unit vector
\begin{equation}
\left\langle \epsilon_{p}\epsilon_{q}\right\rangle _{\chi}=\frac{1}{2}\left(e_{\theta;p}e_{\theta;q}+e_{\phi;p}e_{\phi;q}\right)=\frac{1}{2}\left(\delta_{pq}-e_{r;p}e_{r;q}\right),\label{eq:chi-average}
\end{equation}
which follows from the orthonormality of the unit vectors, Eq.~\eqref{eq:radial unit vector}.
The $\left(\theta,\phi\right)$-average, on the other hand, will be
handled numerically using Lebedev quadrature, \cite{lebedev1975values,lebedev1976quadratures,lebedev1977spherical,lebedev1992quadrature,lebedev1995quadrature,lebedev1999quadrature}
which we in our corresponding non-relativistic work have found to
converge quickly.\cite{List_MP2017}

\subsubsection{Truncated light-matter interaction}

In the generalized velocity representation of Section \ref{subsec:Coulomb gauge}), the rotational average initially reads
\begin{eqnarray}
\left\langle \sigma^{[2n]}\left(\omega\right)\right\rangle _{\theta,\phi,\chi} & = & \frac{\pi\omega}{\varepsilon_{0}\hbar c}\left(\frac{e}{\omega_{fi}}\right)^{2}\sum_{m=0}^{n}\frac{\left(-1\right)^{m}}{\left(n+m\right)!\left(n-m\right)!}\left(2-\delta_{m0}\right)\left(\frac{\omega_{fi}}{c}\right)^{2n}\left\langle \left\langle \epsilon_{p}\epsilon_{q}\right\rangle _{\chi}e_{r;j_{1}}e_{r;j_{2}}\ldots e_{r;j_{2n}}\right\rangle _{\theta,\phi}\nonumber\\
 & \times & \text{Re}\left\{ \langle f|ic\alpha_{p}r_{j_{1}}\ldots r_{j_{n+m}}|i\rangle\langle f|ic\alpha_{q}r_{j_{n+m+1}}\ldots r_{j_{2n}}|i\rangle^{\ast}\right\} f\left(\omega,\omega_{fi},\gamma_{fi}\right),
\end{eqnarray}
whereas in the generalized length representation (multipolar gauge), the corresponding expression is
\begin{eqnarray}
\left\langle \sigma^{[2n]}\left(\omega\right)\right\rangle _{\theta,\phi,\chi} & = & \frac{\pi\omega}{\varepsilon_{0}\hbar c}\sum_{m=0}^{n}\left(-1\right)^{m}\left(2-\delta_{m0}\right)\left(\frac{\omega_{fi}}{c}\right)^{2n}\left\langle \left\langle \epsilon_{p}\epsilon_{q}\right\rangle _{\chi}e_{r;j_{1}}e_{r;j_{2}}\ldots e_{r;j_{2n}}\right\rangle _{\theta,\phi}\nonumber\\
 & \times & \text{Re}\left\{ \langle f|\hat{X}_{j_{1}\ldots j_{n+m};p}^{[n+m]}\left(\omega\right)|i\rangle\langle f|\hat{X}_{j_{n+m+1}\ldots j_{2n};q}^{[n-m]}\left(\omega\right)|i\rangle^{\ast}\right\} f\left(\omega,\omega_{fi},\gamma_{fi}\right).
\end{eqnarray}
 In both cases, the central quantity to evaluate is
\begin{align}
\left\langle \left\langle \epsilon_{p}\epsilon_{q}\right\rangle _{\chi}e_{r;j_{1}}e_{r;j_{2}}\ldots e_{r;j_{2n}}\right\rangle _{\theta,\phi}=\frac{1}{4\pi}\int_{0}^{2\pi}\int_{0}^{\pi}\left\langle \epsilon_{p}\epsilon_{q}\right\rangle _{\chi}e_{r;j_{1}}e_{r;j_{2}}\ldots e_{r;j_{2n}}\sin\theta d\theta d\phi.
\end{align}
 Since the integrand is fully symmetric in indices $\left(j_{1},\ldots j_{2n}\right)$,
we can collect contributions to the three components of the wave unit
vector to give
\begin{align} \label{eq:rotav}
\left\langle \left\langle \epsilon_{p}\epsilon_{q}\right\rangle _{\chi}e_{r;j_{1}}e_{r;j_{2}}\ldots e_{r;j_{2n}}\right\rangle _{\theta,\phi} &=\notag\\
  \frac{1}{8\pi}\int_{0}^{2\pi}\int_{0}^{\pi}&\left(\delta_{pq}-e_{r;p}e_{r;q}\right)e_{r;x}^{i}e_{r;y}^{j}e_{r;z}^{k}\sin\theta d\theta d\phi; \quad i+j+k=2n.
\end{align}
 The calculation of the rotational averages thus hinges on the evaluation
of expressions of the form
\begin{align}
E_{tuv}=\frac{1}{8\pi}\int_{0}^{2\pi}\int_{0}^{\pi}e_{r;x}^{t}e_{r;y}^{u}e_{r;z}^{v}\sin\theta d\theta d\phi=\frac{1}{8\pi}\int_{0}^{2\pi}\cos^{t}\phi\sin^{u}\phi\int_{0}^{\pi}\cos^{v}\theta\sin^{t+u+1}\theta d\theta d\phi.
\end{align}

A computational useful expression is obtained in two steps. First
we use the relations
\begin{eqnarray}
\int_{0}^{\pi}\cos^{p}\theta\sin^{q}\theta d\theta & = & \left(1+\left(-1\right)^{p}\right)\int_{0}^{\pi/2}\cos^{p}\theta\sin^{q}\theta d\theta\\
\int_{0}^{2\pi}\cos^{p}\phi\sin^{q}\phi d\phi & = & \left(1+\left(-1\right)^{p}\right)\left(1+\left(-1\right)^{q}\right)\int_{0}^{\pi/2}\cos^{p}\phi\sin^{q}d\phi,
\end{eqnarray}
 to reduce the angular integration to the $(+,+,+)$ octant of Euclidean
space
\begin{align}
E_{tuv}=\frac{1}{8\pi}\left[1+\left(-1\right)^{t}\right]\left[1+\left(-1\right)^{u}\right]\left[1+\left(-1\right)^{v}\right]\int_{0}^{\pi/2}\int_{0}^{\pi/2}e_{r;x}^{t}e_{r;y}^{u}e_{r;z}^{v}\sin\theta d\theta d\phi.
\end{align}
 This provides a powerful selection rule, showing that the expression
$E_{tuv}$ is zero unless all integer exponents $t$, $u$ and $v$
are even. In passing, we note that the selection rule is the same for
both terms appearing in Eq.~\eqref{eq:rotav} for $p=q$. Second, we use
the integral representation (see Appendix \ref{sec:The-trivariate-beta})
\begin{equation}
B\left(a,b,c\right)=\frac{\Gamma\left(a\right)\Gamma\left(b\right)\Gamma\left(c\right)}{\Gamma\left(a+b+c\right)}=4\int_{0}^{\pi/2}\int_{0}^{\pi/2}e_{r;x}^{2a-1}e_{r;y}^{2b-1}e_{r;z}^{2c-1}\sin\theta d\theta d\phi,\label{eq:trivariate beta function}
\end{equation}
to express the rotational average in terms of the trivariate beta
function $B\left(a,b,c\right)$
\begin{equation}
E_{tuv}=\frac{1}{32\pi}\left[1+\left(-1\right)^{t}\right]\left[1+\left(-1\right)^{u}\right]\left[1+\left(-1\right)^{v}\right]B\left(\frac{t+1}{2},\frac{u+1}{2},\frac{v+1}{2}\right).\label{eq:rotav_tuv}
\end{equation}
 The final result is thereby
\begin{align}
E_{tuv}=\begin{cases}
\frac{\left(t-1\right)!!\left(u-1\right)!!\left(v-1\right)!!}{2\left(t+u+v+1\right)!!} & t,u,v\text{ even}\\
0 & \text{otherwise}
\end{cases},
\end{align}
 where we have used the identity
\begin{align}
\Gamma\left(\frac{t+1}{2}\right)=\left(t-1\right)!!\sqrt{\frac{\pi}{2^{t}}}
\end{align}
 for the evaluation of the trivariate beta function.

Our approach is different from the conventional approach
to rotational averages using linear combinations of fundamental Cartesian
isotropic tensors.\cite{Kearsley_Fong_1975,Andrews_Thirunamachandran_JCP1977,Andrews_Ghoul_JPhysA1981,Andrews_Blake_JPhysA1989,Friese_JCP2014,Ee_EuJPhys2017}
The fundamental Cartesian isotropic tensors of even rank are given
by products of Kronecker deltas $\delta_{ij}$, whereas an additional
Levi-Civita symbol $\epsilon_{ijk}$ appears at odd rank.\cite{weyl1939groups,hodge1961isotropic,Jeffreys_1973}
For instance, connecting to the notation of Barron,\cite{barron2004molecular}
the rotational average appearing in the second-order contribution
$\sigma^{\left[2\right]}$ to the absorption cross section is
\begin{align}
\left\langle \left\langle \epsilon_{\alpha}\epsilon_{\beta}\right\rangle _{\chi}e_{r;\gamma}e_{r;\delta}\right\rangle _{\theta,\phi}=\left\langle i_{\alpha}i_{\beta}k_{\gamma}k_{\delta}\right\rangle =\frac{1}{30}\left(4\delta_{\alpha\beta}\delta_{\gamma\delta}-\delta_{\alpha\gamma}\delta_{\beta\delta}-\delta_{\alpha\delta}\delta_{\beta\delta}\right).
\end{align}
The established procedure for generating a suitable linearly independent
set of fundamental Cartesian isotropic tensors involves the construction
of standard tableaux from Young diagrams.\cite{Smith_Tensor1968,Andrews_Thirunamachandran_JCP1977}
For even rank, one can connect to our approach from the observation
that the integer exponents $t$, $u$ and $v$ in the expression for $E_{tuv}$
(Eq.~\eqref{eq:rotav_tuv}) must all be even. This implies a pairing of
indices, which can be expressed through strings of Kronecker deltas. Simple
combinatorics suggests that the possible number of pairings of $2n$
indices and thus the number of fundamental Cartesian isotropic tensors
of even rank $2n$ is $(2n-1)!!$. However, starting at rank 8 linear
dependencies (syzygies) occur, e.g.\cite{Rivlin1955,Kearsley_Fong_1975,Andrews_Ghoul_JPhysA1981}
\begin{equation}
\left|\begin{array}{cccc}
\delta_{i_{1}i_{5}} & \delta_{i_{1}i_{6}} & \delta_{i_{1}i_{7}} & \delta_{i_{1}i_{8}}\\
\delta_{i_{2}i_{5}} & \delta_{i_{2}i_{6}} & \delta_{i_{2}i_{7}} & \delta_{i_{2}i_{8}}\\
\delta_{i_{3}i_{5}} & \delta_{i_{3}i_{6}} & \delta_{i_{3}i_{7}} & \delta_{i_{3}i_{8}}\\
\delta_{i_{4}i_{5}} & \delta_{i_{4}i_{6}} & \delta_{i_{4}i_{7}} & \delta_{i_{4}i_{8}}
\end{array}\right|=0,
\end{equation}
which requires proper handling. In fact, the number of linearly independent
fundamental Cartesian isotropic tensors of a given rank is given by
Motzkin sum numbers\cite{Motzkin} which for rank 8 is 91 rather than
105 suggested by the double factorial derived for even rank above.
Such considerations are not needed in the present approach which in
addition is well-suited for computer implementation.

\section{Computational Details}
Unless otherwise stated the calculated results presented in this paper have been obtained by time-dependent density functional theory (TD-DFT) calculations, based on the Dirac--Coulomb Hamiltonian and within the
restricted excitation window (REW) approach\cite{Stener_CPL2003,South_PCCP2016} using the PBE0\cite{perdew1996generalized,adamo1999toward} exchange-correlation functional and the dyall.ae3z basis sets.\cite{dyall2009relativistic,dyall2012core} The small component basis sets were generated according to the condition of restricted kinetic balance, and
the $(\text{SS}|\text{SS})$ integrals are replaced by an interatomic SS correction.\cite{visscher1997approximate} A Gaussian model was employed for the nuclear charge distribution.\cite{Visscher_Dyall_nuc} A 86-point Lebedev grid ($L_\text{max}=12$) was used for the isotropic averaging of the oscillator strengths based on the full light-matter interaction operator. The gauge origin was placed in the center-of-mass and spatial symmetry was invoked in all cases except for the gauge-origin dependence calculations. 

The geometry of \ce{TiCl4} was taken from  Ref.~\citenum{bernadotte_JCP2012} where it was obtained using the BP86 exchange-correlation
functional\cite{Becke_PRA1988,Perdew_PRB1986,*Perdew_PRB1986err} and the TZP basis set.\cite{ADF_basis_2003} 
To enable a direct comparison to previous work,\cite{lestrange_JCP2015} additional results on \ce{TiCl4} have been obtained using the non-relativistic L\'{e}vy-Leblond Hamiltonian\cite{levy1967nonrelativistic} employing a point-nucleus model and the 6-31+G* basis set,\cite{hehre1972self,hariharan1973influence,francl1982self,rassolov19986} the latter as implemented in the Gaussian16 package.\cite{g16}

To study the apparent divergences of oscillator strengths for core excitations using truncated interaction, we carried out time-dependent Hartree--Fock (TD-HF) calculations of $ns_{1/2}\rightarrow 7p_{1/2}$ excitations of the radium atom. In these calculations integral screening was turned off and the $(\text{SS}|\text{SS})$ integrals included.

Unless otherwise stated the data reported in this paper have been obtained with a development version of the \textsc{Dirac} electronic structure code\cite{DIRAC19} 
(Tables \ref{tab:TICL4}--\ref{tab:TICL4_gaugetest2}, Figure~\ref{fig:Ra}: revision \texttt{52c65be}; Table \ref{tab:Ra_vel}; Figure \ref{fig:Ra_convergence_test}: revision \texttt{5a7d81c}).

\section{Results and Discussion\label{subsec:results_and_discussion}}
In this section, we demonstrate our implementation and study the behavior of the three presented schemes to go beyond the ED approximation. First, we consider the \ce{Cl} \textit{K}-edge in \ce{TiCl4}, representing a 
case where there is no natural choice of gauge origin. It has previously been studied in the context of non-dipolar effects in linear X-ray absorption using low-order multipole expansions. In particular, it was used to demonstrate the appearance of negative oscillator strengths\cite{lestrange_JCP2015} upon truncation of the light-matter interaction in the generalized velocity representation in a non-relativistic framework.\cite{bernadotte_JCP2012} Below, we will revisit this case. We further study numerically the gauge-origin dependence of the three schemes in the case of soft X-ray absorption. 
We then turn to their performance across the spectral range, including hard X-rays,  by considering atomic valence and core transitions in the radium atom. 
Given its high nuclear charge, radium shows strong relativistic effects both in the core and valence, and it is therefore a good example for comparing oscillator strengths within and beyond the ED approximation in a relativistic framework.

\subsection{Cl \textit{K}-edge absorption of \ce{TiCl4}}
Ligand \textit{K}-edge absorption spectroscopy supposedly provides direct information on the covalency of metal--ligand bonds due to the admixture of the ligand $p$-orbitals with the metal $d$-orbitals.\cite{glaser2000ligand,solomon2005ligand} The \ce{Cl}  \textit{K}-edge absorption of \ce{TiCl4}  has been studied both experimentally and also theoretically within and beyond the ED approximation using truncated multipole-expanded expressions.  Its experimental spectrum features a broad pre-edge peak that require a two-peak fit (in toluene: at 2821.58 and 2822.32 eV with an approximate intensity ratio of 0.84).\cite{debeer2005metal} In T$_d$ symmetry, the five $3d$-orbitals of Ti belong to the $e$ and $t_2$ irreducible representations, and the pre-edge bands can be assigned to excitations from the $a_1$ and $t_2$ Cl $1s$-orbitals into the $e$ and $t_2$ sets of $3d$-orbitals on Ti, respectively. Here, we focus on the eight lowest-lying transitions ($a_1, t_2\rightarrow e$) which gives rise to three degenerate sets ($E$, $T_1$ and $T_2$) of which the latter is ED allowed. 

\subsubsection{Full vs.~truncated light-matter interaction}
Table \ref{tab:TICL4} collects the isotropically averaged oscillator strengths for the pre-edge transitions computed in 4-component relativistic and non-relativistic frameworks with the full light-matter interaction operator as well as accumulated to increasing orders (up to 12th order) in the wave vector within Coulomb gauge (velocity representation) and multipolar gauge (length representation). First, we note that the trends are similar across the considered basis sets and Hamiltonians. In line with the results of Lestrange \textit{et al.},\cite{lestrange_JCP2015} we find negative oscillator strengths at second order for the $^1T_2$ excitations in both length and velocity representation. The same issue appears for the $^1T_1$ and $^1E$ sets, but at fourth order. 
As discussed previously,\cite{lestrange_JCP2015,Sorensen_MP2016} this behavior is expected when the cross terms involving the lower-order moments to $f^{[n]}$ dominate the diagonal contributions. As evident from the underlying contributions given in Table \ref{tab:TICL4_2}, the multipole expansions are alternating, and beyond fourth order, the correction is reduced at each order. 
Indeed, the expansions converge to the full expression at about 12th order irrespective of the employed basis set. 
For the dipole-allowed $^1T_2$ set, the correction introduced by non-dipolar effects is significant, reducing the oscillator strength by a factor of $\sim$5. As seen from the comparison of the ED and full (BED) oscillator strengths summed over the three sets of transitions, included in Table \ref{tab:TICL4}, the implication of going beyond the ED approximation is a redistribution of intensity among transitions.
In particular, the ED forbidden $^1T_1$ and $^1E$ transitions gain intensity beyond that of the $T_2$ set.  
We note, however, that this intensity redistribution has no consequence for the absorption band because of the near-degeneracy of the electronic transitions.

\begin{table}[h!]
	\scriptsize
	\caption{Comparison of isotropically averaged oscillator strengths for \ce{Cl} $1s\rightarrow$\ce{Ti} $3d$ transitions of \ce{TiCl4} for the full semi-classical interaction operator and accumulated to various orders, as indicated by the superscripted number in parenthesis, within multipolar gauge (lr: length representation) and Coulomb gauge (vr: velocity representation), computed at the 4c-TD-PBE0 and L\'{e}vy-Leblond (LL) level of theory with different basis sets. Contributions from degenerate states have been summed. A 86-point ($L_\text{max}=12$) Lebedev grid was used to obtain the isotropically averaged full BED oscillator strengths. The gauge origin is placed on the Ti atom.}
	\begin{threeparttable}
		\begin{tabular*}{\textwidth}{@{\extracolsep{\fill}}ccccccccccc}
			\hline \hline\noalign{\smallskip}
			Final state & $\Delta E$ (eV) & gauge & $10^{3}f^{(\rightarrow0)}$ & $10^{3}f^{(\rightarrow2)}$ & $10^{3}f^{(\rightarrow4)}$ &$10^{3}f^{(\rightarrow6)}$ & $10^{3}f^{(\rightarrow8)}$ & $10^{3}f^{(\rightarrow10)}$ & $10^{3}f^{(\rightarrow12)}$   &$10^{3}f_{\text{full}}$\\
			\hline\noalign{\smallskip}
			\multicolumn{11}{c}{6-31+G* -- LL}\\
			\hline\noalign{\smallskip}
			\multirow{2}{*}{$^{1}T_{1}$} & \multirow{2}{*}{2763.004298} & \text{lr} & 0.000 & 16.616 & -6.599 & 7.867 & 2.748 & 3.913 & 3.730 & \multirow{2}{*}{3.730} \\
			& & \text{vr} & 0.000 & 16.616 & -6.598 & 7.877 & 2.715 & 3.900 & 3.709 &  \\
			& 2763.004474\tnote{\textit{a}}& \text{vr} & 0.000 & 16.62 &- &- &-& -& -&  \\
			\multirow{2}{*}{$^{1}E$}     & \multirow{2}{*}{2763.004339} & \text{lr} & 0.000 & 6.762 & -1.188 & 3.360 & 1.814 & 2.164 & 2.112 & \multirow{2}{*}{2.096} \\
			&  & \text{vr} & 0.000 & 6.640 & -1.109 & 3.288 & 1.816 & 2.141 & 2.090 &  \\
			& 2763.004515\tnote{\textit{a}} & \text{vr}&  0.000 &  6.64 & - &- &-& -& -&  \\
			\multirow{2}{*}{$^{1}T_{2}$} & \multirow{2}{*}{2763.004306} & \text{lr} & 7.434 & -16.230 & 15.073 & -3.955 & 2.669 & 1.198 & 1.408 &\multirow{2}{*}{1.396}\\
			&  & \text{vr} & 7.246 & -16.033 & 14.988 & -3.943 & 2.690 & 1.180 & 1.422 &\\
			& 2763.004482\tnote{\textit{a}} & \text{vr}&  7.44 & -15.84 & - &- &-& -& -&  \\
			\cline{2-11}	\noalign{\smallskip}
			& \multirow{2}{*}{\text{Sum}}& \text{lr} &  7.434 & 7.147 &7.286 & 7.273 & 7.231 &	7.275 &	7.249& \multirow{2}{*}{7.222}\\
			& & \text{vr}& 	7.246 &	7.222 & 7.221 & 7.222 & 7.221 & 7.222& 7.221& \\
			\multicolumn{11}{c}{dyall.ae3z -- LL}\\
			\hline\noalign{\smallskip}
			\multirow{2}{*}{$^{1}T_{1}$} & \multirow{2}{*}{2762.623981} & \text{lr} & {0.000} & 17.964  & -7.142 & 8.505 & 2.959& 4.230& 4.026& \multirow{2}{*}{4.040}\\
			&  & \text{vr} & {0.000} & 17.964  & -7.166 & 8.504 & 2.948& 4.222& 4.017& \\
			\multirow{2}{*}{$^{1}E$}     & \multirow{2}{*}{2762.623987} & \text{lr} & 0.000 & 7.192 & -1.179  & 3.553 & 1.976 & 2.322 & 2.269 & \multirow{2}{*}{2.267}\\
			&  & \text{vr} & 0.000 & 7.151 & -1.161  & 3.536 & 1.971 & 2.314 & 2.261 & \\
			\multirow{2}{*}{$^{1}T_{2}$} & \multirow{2}{*}{2762.623987} & \text{lr}  & 7.880 & -17.335 & 16.149 & -4.230 & 2.893& 1.276&1.533 & \multirow{2}{*}{1.503}\\
			&  & \text{vr}  & 7.836 & -17.305 & 16.137 & -4.232 & 2.890& 1.273 & 1.531 & \\
			\cline{2-11}	\noalign{\smallskip}
			& \multirow{2}{*}{\text{Sum}}& \text{lr}& 7.880 & 7.821 & 7.828 & 7.829 &7.828 &7.828 & 7.828 & \multirow{2}{*}{7.809}\\
			& & \text{vr} &  7.836 & 7.809& 7.809 & 7.809 & 7.809 & 7.809&	7.809& \\
			\multicolumn{11}{c}{dyall.ae3z -- 4c}\\
			\hline\noalign{\smallskip}
			\multirow{2}{*}{$^{1}T_{1}$} & \multirow{2}{*}{2773.351719} & \text{lr} & 0.000 & 17.976  & -7.344 & 8.560& 2.879 & 4.191 &3.979 & \multirow{2}{*}{3.993}\\
			& & \text{vr} & 0.000 & 17.976  & -7.372 & 8.561& 2.866 & 4.183 &3.970 & \\
			\multirow{2}{*}{$^{1}E$}     & \multirow{2}{*}{2773.351723} & \text{lr} & 0.000 & 7.199 & -1.251 & 3.569 & 1.945 & 2.308 &2.249 &  \multirow{2}{*}{2.248} \\
			& & \text{vr}& 0.000 & 7.156 & -1.229 & 3.548 & 1.943 & 2.298 &2.242 &  \\
			\multirow{2}{*}{$^{1}T_{2}$} & \multirow{2}{*}{2773.351725} & \text{lr} & 7.825 & -17.413 & 16.369& -4.360 & 2.948& 1.272 & 1.543 & \multirow{2}{*}{1.510}\\
			&  & \text{vr} & 7.781 & -17.380 & 16.353 & -4.358 & 2.942& 1.271 & 1.540 & \\
			\cline{2-11}	\noalign{\smallskip}
			&\multirow{2}{*}{\text{Sum}} & \text{lr}& 7.825 & 7.763 & 7.775 & 7.769 & 7.772 & 7.772 & 7.771 & \multirow{2}{*}{7.752}\\
			& & \text{vr} &  7.781 &7.752 & 7.752 &7.752 &7.752 &7.752 &7.752 & \\
			\noalign{\smallskip}\hline \hline
		\end{tabular*}
		\begin{tablenotes}
			\item[a] Data in row taken from Ref.~\citenum{lestrange_JCP2015}.
		\end{tablenotes}
	\end{threeparttable}
	\label{tab:TICL4}
\end{table}

\begin{table}[h!]
	\scriptsize
	\caption{Comparison of isotropically averaged oscillator strengths for Cl $1s\rightarrow$Ti $3d$ transitions of \ce{TiCl4} for the full BED operator and at various orders within multipolar gauge (lr: length representation) and Coulomb gauge (vr: velocity representation) gauges as computed at the 4c-TD-PBE0 and L\'{e}vy-Leblond (LL) level of theory with different basis sets. Numbers in parentheses are exponents of 10. Contributions from degenerate states have been summed. A 86-point ($L_\text{max}=12$) Lebedev grid was used to obtain the isotropically averaged full BED oscillator strengths. The gauge origin is placed on the Ti atom.}
	\begin{tabular*}{\textwidth}{@{\extracolsep{\fill}}ccccccccccc}
		\hline \hline\noalign{\smallskip}
		Final state & $\Delta E$ (eV) & gauge & $f^{[0]}$ & $f^{[2]}$ & $f^{[4]}$ &$f^{[6]}$ & $f^{[8]}$ & $f^{[10]}$ & $f^{[12]}$   &$f_{\text{full}}$\\
		\hline\noalign{\smallskip}
		\multicolumn{11}{c}{6-31+G* -- LL}\\
		\hline\noalign{\smallskip}
		\multirow{2}{*}{$^{1}T_{1}$} & \multirow{2}{*}{2763.004298} & \text{lr} & 0.000 & 1.662(-02) & -2.321(-02) & 1.447(-02) & -5.120(-03) & 1.166(-03) & -1.837(-04) & \multirow{2}{*}{3.730(-03)} \\
		& & \text{vr} & 0.000 & 1.662(-02) & -2.327(-02) & 1.453(-02) & -5.162(-03) & 1.186(-03) & -1.910(-04) &  \\
		\multirow{2}{*}{$^{1}E$}     & \multirow{2}{*}{2763.004339} & \text{lr} & 0.000 & 6.762(-03) & -7.949(-03)& 4.548(-03) &-1.545(-03) & 3.493(-04) & -5.194(-05) & \multirow{2}{*}{2.096(-03)} \\
		&  & \text{vr} & 0.000 & 6.640(-03) & -7.749(-03) & 4.397(-03) & -1.471(-03) & 3.245(-04) & -5.069(-05) &  \\
		\multirow{2}{*}{$^{1}T_{2}$} & \multirow{2}{*}{2763.004306} & \text{lr} & 7.434(-03) & -2.366(-02) & 3.130(-02) & -1.903(-02) & 6.624(-03)& -1.471(-03) & 2.098(-04) &\multirow{2}{*}{1.396(-03)}\\
		&  & \text{vr} & 7.246(-03) & -2.328(-02) & 3.102(-02) & -1.893(-02) & 6.633(-03) & -1.509(-03) & 2.413(-04) &\\[0.1in]
		\multicolumn{11}{c}{dyall.ae3z -- LL}\\
		\hline\noalign{\smallskip}
		\multirow{2}{*}{$^{1}T_{1}$} & \multirow{2}{*}{2762.623981} & \text{lr} & {0.000} & 1.796(-02)  & -2.511(-02) & 1.565(-02) & -5.546(-03) & 1.271(-03) & -2.040(-04)& \multirow{2}{*}{4.040(-03)}\\
		&  & \text{vr} & {0.000} & 1.796(-02)  & -2.513(-02) & 1.567(-02) & -5.556(-03) & 1.274(-03)& -2.045(-04)& \\
		\multirow{2}{*}{$^{1}E$}     & \multirow{2}{*}{2762.623987} & \text{lr} & 0.000 & 7.192(-03) & -8.372(-03)  & 4.733(-03) & -1.577(-03) & 3.460(-04) & -5.346(-05) & \multirow{2}{*}{2.267(-03)}\\
		&  & \text{vr} & 0.000 & 7.151(-03) & -8.312(-03)  & 4.698(-03) & -1.566(-03) & 3.439(-04) & -5.348(-05) & \\
		\multirow{2}{*}{$^{1}T_{2}$} & \multirow{2}{*}{2762.623987} & \text{lr}  & 7.880(-03) & -2.521(-02) & 3.348(-02) & -2.038(-02) & 7.122(-03)& -1.617(-03)&2.574(-04) & \multirow{2}{*}{1.503(-03)}\\
		&  & \text{vr}  & 7.836(-03) & -2.514(-02) & 3.344(-02) & -2.037(-02) & 7.122(-03)& -1.618(-03) & 2.580(-04) & \\[0.1in]
		\multicolumn{11}{c}{dyall.ae3z -- 4c}\\
		\hline\noalign{\smallskip}
		\multirow{2}{*}{$^{1}T_{1}$} & \multirow{2}{*}{2773.351719} & \text{lr} & 0.000 & 1.798(-02)  & -2.532(-02) & 1.590(-02)& -5.682(-03) & 1.312(-03) & -2.125(-04) & \multirow{2}{*}{3.993(-03)}\\
		& & \text{vr} & 0.000 & 1.798(-02)  & -2.535(-02) & 1.593(-02) & -5.695(-03) & 1.316(-03) & -2.131(-04) & \\
		\multirow{2}{*}{$^{1}E$}     & \multirow{2}{*}{2773.351723} & \text{lr} & 0.000 & 7.199(-03) & -8.450(-03) & 4.819(-03) & -1.623(-03) & 3.627(-04) & -5.937(-05) &  \multirow{2}{*}{2.248(-03)} \\
		& & \text{vr}& 0.000 & 7.156(-03) & -8.385(-03) & 4.778(-03) & -1.606(-03) & 3.555(-04) & -5.576(-05) &  \\
		\multirow{2}{*}{$^{1}T_{2}$} & \multirow{2}{*}{2773.351725} & \text{lr} & 7.825(-03) & -2.524(-02) & 3.378(-02)& -2.073(-02) & 7.308(-03)& -1.675(-03) & 2.706(-04) & \multirow{2}{*}{1.510(-03)}\\
		&  & \text{vr} & 7.781(-03) & -2.516(-02) & 3.373(-02) & -2.071(-02) & 7.301(-03) & -1.672(-03) & 2.689(-04) & \\
		\noalign{\smallskip}\hline \hline
	\end{tabular*}
	\label{tab:TICL4_2}
\end{table}

\subsubsection{Origin-dependence}\label{sec:origindep}
The above results were computed with the gauge-origin placed at the Ti atom. We now proceed to a numerical evaluation of their dependency on the gauge origin ($\mathbf{O}+\mathbf{a}$). 
As discussed above, the formulations based on the full semi-classical interaction operator and truncated interaction in the velocity representation  are formally gauge invariant. In practical calculations, however,
as discussed in Section \ref{subsec:Coulomb gauge}, invariance in the latter case relies on the accurate cancellation of lower-order contributions multiplied with powers $(\mathbf{k}\cdot\mathbf{a})$, where $\mathbf{a}$ is the displacement. In contrast, as discussed in Section \ref{subsec:mg}, in the multipolar gauge formal gauge-origin invariance appears to only be achieved in the practically unreachable limit of the complete expansion of the fields.

Table \ref{tab:TICL4_gaugetest2} collects the total isotropic oscillator strength for the dipole-allowed $^1T_2$ set for each of the three schemes for going beyond the ED approximation using different choices for the gauge origin. As expected, the results for the full light-matter interaction operator remain unchanged, providing a numerical verification of its gauge-origin invariance. The same is true for the oscillator strengths in the generalized velocity representation. However, numerical noise from the cancellation of many terms in powers of the displacement becomes apparent at large displacements. For a displacement of 100 $a_0$, instabilities start to appear at 10th order, and at 12th order, the oscillator strength exceeds the full result by one order of magnitude. This will be further discussed in Section \ref{subsec:Ra_truncated}. The oscillator strengths in the multipolar gauge already at second order differ significantly upon shifting the origin from the Ti atom.

\begin{table}[h!]
	\scriptsize
	\caption{Gauge-origin dependency of the isotropically averaged oscillator strengths for the $ ^{1}T_{2}$ set of \ce{Cl} $1s\rightarrow$\ce{Ti} $3d$ transitions of \ce{TiCl4} for the full semi-classical light-matter interaction operator and accumulated to various orders within multipolar gauge (lr: length representation) and Coulomb gauge (vr: velocity representation), computed at the 4c-TD-PBE0 level of theory and the dyall.ae3z basis set. Numbers in parentheses are exponents of 10. At this level the excitation energy is calculated as 2773.351145 eV. Contributions from the degenerate set have been summed. A 86-point ($L_\text{max}=12$) Lebedev grid was used to obtain the isotropically averaged full BED oscillator strengths. The gauge origin is shifted along the $x$-axis ($d_x$) where $d_x=0.0\ a_0$ corresponds to gauge-origin in the Ti atom.} 
	\begin{tabular*}{\textwidth}{@{\extracolsep{\fill}}cccccccccc}
		\hline \hline\noalign{\smallskip}
		$d_x$ ($a_0$) &  gauge & $f^{(\rightarrow 0)}$ & $f^{(\rightarrow 2)}$ & $f^{(\rightarrow 4)}$ &$f^{(\rightarrow 6)}$ & $f^{(\rightarrow 8)}$ & $f^{(\rightarrow 10)}$ & $f^{(\rightarrow 12)}$   &$f_{\text{full}}$\\
		\hline\noalign{\smallskip}
\multirow{2}{*}{0}    & \text{lr} & 7.825(-03) & -1.741(-02)&	1.637(-02) & -4.360(-03) & 2.948(-03) &  1.272(-03) & 1.543(-03) &\multirow{2}{*}{1.510(-03)}\\
		      & \text{vr} & 7.781(-03) & -1.738(-02) &	1.635(-02) & -4.358(-03) & 2.943(-03) &  1.271(-03) & 1.540(-03) &\\[0.1in]
\multirow{2}{*}{10.0} & \text{lr} & 7.825(-03) &	-1.755(-02) &	1.670(-02) & -4.738(-03) & 3.238(-03) &  1.097(-03) & 1.670(-03) & \multirow{2}{*}{1.510(-03)}\\
		      & \text{vr} & 7.781(-03) &	-1.738(-02) &	1.635(-02) & -4.358(-03) & 2.943(-03) &  1.271(-03) & 1.540(-03) & \\[0.1in]
\multirow{2}{*}{50.0} & \text{lr} & 7.825(-03) & -2.045(-02) &	1.422(-01) & -2.951(+00) & 4.429(+01) & -5.055(+02) & 6.495(+03) & \multirow{2}{*}{1.510(-03)}\\
		      & \text{vr} & 7.781(-03) &	-1.738(-02) &	1.635(-02) & -4.358(-03) & 2.943(-03) &  1.271(-03) & 1.546(-03) & \\[0.1in]
\multirow{2}{*}{100.0}& \text{lr} & 7.825(-03) &	-3.148(-02) &    2.398(+00) & -2.223(+02) & 1.343(+04) & -6.178(+05) & 3.198(+07) & \multirow{2}{*}{1.510(-03)}\\
		      & \text{vr} & 7.781(-03) & -1.738(-02) &	1.635(-02) & -4.358(-03) & 2.943(-03) &  1.021(-03) & 4.785(-02) & \\
		\noalign{\smallskip}\hline \hline
	\end{tabular*}
	\label{tab:TICL4_gaugetest2}
\end{table}

\subsection{Radium}

\subsubsection{Full light-matter interaction}
In the valence region, the influence of non-dipolar effects is expected to be small except for ED forbidden transitions. Based on our previous study in a non-relativistic framework, we expect the effect on dipole-allowed core excitations to be modest (${\sim}10\%$) as a result of the compactness of the core hole.\cite{List_JCP2015}  

\begin{figure}[ht!]
	\centering
	\includegraphics[width=0.6\columnwidth]{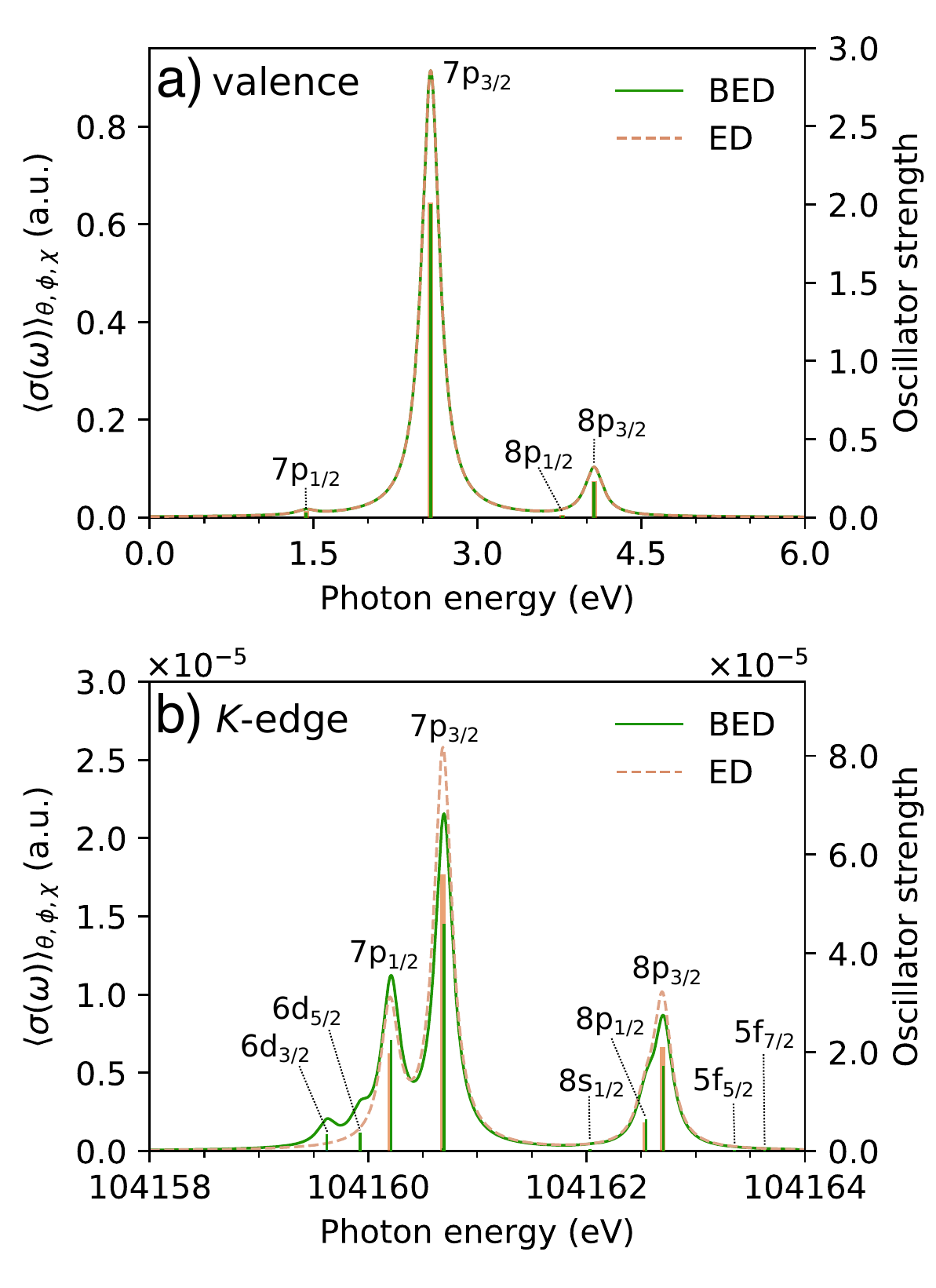}
	\caption{Non-dipolar effects on electronic absorption of radium: (a) the valence and (b) \textit{K}-edge spectra for Ra within and beyond the ED approximation (ED and BED, respectively) at the 4c-TD-PBE0/dyall.ae3z level of theory, using Coulomb gauge (velocity representation) for the former and the full interaction operator in Eq.~\eqref{eq:avg BED osc} and an 86-point ($L_\text{max}=12$) Lebedev grid for the latter. The labels indicate the character of the receiving orbital. Note the differences in scales on the axes in the valence and X-ray region. Oscillator strengths are summed over contributions from transitions within each degenerate (same $\Delta J$ components) and near-degenerate (different $\Delta J$ components) set, and the sticks have been convoluted with a Lorentzian lifetime broadening of 1000 cm$^{-1}$. The experimental $1s$ ionization energy is $103922\pm7.2$ eV.\cite{bearden1967reevaluation}
	}
	\label{fig:Ra}%
\end{figure}

Figure \ref{fig:Ra} shows the valence and \textit{K}-edge spectra of Ra within and beyond the ED approximation, the latter computed with the full light-matter interaction operator. 
Expectedly, all ED forbidden transitions, except for excitations associated with $\Delta J=0$ change in total angular momentum quantum number, gain intensity upon going beyond the ED approximation. In the valence region, however, they remain several orders of magnitude smaller than the ED counterparts, such that ED and BED spectra are essentially identical.
In the X-ray region, the main contributions from the $1s_{1/2}\rightarrow 6d$ manifold corresponds to $\Delta J=2$ transitions, while the ED allowed $\Delta J=1$ transitions dominate for the $1s_{1/2}\rightarrow 7/8p$ manifold. Note that the small energy differences between different  $\Delta J$ components in a given set makes them indiscernible in the spectrum, and we have therefore combined their oscillator strengths in Figure \ref{fig:Ra}. Upon inclusion of non-dipolar effects, intensity is primarily redistributed from the $1s_{1/2}\rightarrow 7/8p_{3/2}$ sets (a ${\sim}20$\% reduction compared to ${\sim}13$\% for the $1s_{1/2}\rightarrow 7/8p_{1/2}$ excitations) to the $6d$ transitions.

\subsubsection{Truncated light-matter interaction} \label{subsec:Ra_truncated}
When carrying out equivalent calculations using the truncated light-matter interaction formulations, both in the velocity and the length representation, nonsensical results were obtained. Rather than reporting
these numbers, we shall illustrate and analyze this behavior using a simpler computational setup. Table \ref{tab:Ra_vel} reports anisotropic oscillator strengths for radium $ns_{1/2}\rightarrow 7p_{1/2}\ (n=1,..,7)$ excitations at various orders in the generalized velocity representation as well as obtained using the full light-matter interaction. The orbital rotation operator, Eq.~\eqref{eq:orbrot}, is restricted to
the $ns_{1/2}$ and the $7p_{1/2}$ orbitals of the selected excitation, and we only report results for the $B_{1u}$ irreducible representation of the $D_{2h}$ point group. To avoid issues of numerical integration we have performed TD-HF rather than TD-DFT calculations. Furthermore, to avoid possible numerical noise due to rotational averaging, we have choosen an oriented experiment, with the wave and polarization vectors oriented along the $y$- and $z$-axes, respectively.

\begin{sidewaystable*}[!ht]
	\scriptsize
	\caption{Anisotropic oscillator strengths for radium $ns_{1/2}\rightarrow 7p_{1/2}$ excitations at various orders in Coulomb gauge (velocity representation) as well as with the full semi-classical light-matter interaction operator at the 4c-TD-HF/dyall.ae3z level of theory. Numbers in parentheses are exponents of 10. For each excitation, the second line contains the accumulated oscillator strength (ac) to given order. The wave $\mathbf{k}$ and polarization $\boldsymbol{\epsilon}$ vectors were oriented along the $y$- and $z$-axes, respectively. }
	
	\begin{tabular*}{\textwidth}{@{\extracolsep{\fill}}cccccccccccc}
		
		\hline \hline\noalign{\smallskip}
		$n$& $\Delta E$ (eV) & $\omega /c$ (a.u.)  & &   0 &  2 & 4 &6 & 8 & 10&12&$f_{\text{full}}$\\
		\hline\noalign{\smallskip}
		\multirow{2}{*}{7}  &   \multirow{2}{*}{1.82142}       & \multirow{2}{*}{4.88(-04)}   &&6.664(-01) &  9.370(-07) & -2.335(-12) &  6.710(-19) &  2.819(-24) & -5.422(-30) &  5.928(-36) & \\
		&                 &           & ac & 6.664(-01) &6.664(-01) & 6.664(-01) &  6.664(-01) &  6.664(-01) &  6.664(-01) & 6.664(-01) & 6.664(-01) \\
				\hline\noalign{\smallskip}
		\multirow{2}{*}{6}  &  \multirow{2}{*}{41.04937}      & \multirow{2}{*}{1.10(-02)} && 9.850(-05) &  8.799(-08) &  1.253(-11) &  4.800(-15) & -9.481(-19) &  1.982(-22) & -4.937(-26) &  \\
		&                 &         &ac  & 9.850(-01) &  9.859(-05) &  9.859(-05) &  9.859(-05) &  9.859(-05) &  9.859(-05) &  9.859(-05) & 9.841(-05)\\
				\hline\noalign{\smallskip}
		\multirow{2}{*}{5}  & \multirow{2}{*}{269.05081}      & \multirow{2}{*}{7.22(-02)} && 3.947(-04) & -3.662(-07) & -1.568(-10) &  6.514(-14) &  4.330(-15) & -7.432(-17) &  9.082(-19) &  \\
		&                 &           & ac&  3.947(-04)& 3.943(-04) & 3.943(-04) &  3.943(-04) & 3.943(-04) &  3.943(-04) & 3.943(-04) & 3.943(-04)\\
				\hline\noalign{\smallskip}
		\multirow{2}{*}{4}  & \multirow{2}{*}{1240.4258}       & \multirow{2}{*}{3.33(-01)} & & 1.754(-04) & -3.161(-07) & -1.242(-08) &  1.028(-08) & -5.164(-09) &  1.683(-09) &  3.974(-10) &  \\
		&                 &           & ac&1.754(-04) & 1.751(-04) &  1.751(-04) & 1.751(-04) &1.751(-04)& 1.751(-04)&  1.751(-04) &1.751(-04) \\
				\hline\noalign{\smallskip}
		\multirow{2}{*}{3}  & \multirow{2}{*}{4888.5865}       & \multirow{2}{*}{1.31(+00)} & & 6.018(-05) &  2.170(-08) & -3.342(-06) &  4.760(-05) & -3.680(-04) &  1.849(-03) & -6.737(-03) & \\
		&                 &           &ac& 6.018(-05) &6.020(-05) & 5.686(-05) & 1.045(-04) & -2.635(-04) &  1.585(-03) & -5.152(-03) & 6.011(-05) \\
				\hline\noalign{\smallskip}
		\multirow{2}{*}{2}  & \multirow{2}{*}{19398.588}       & \multirow{2}{*}{5.20(+00)} & & 1.784(-05) &  5.379(-07) & -1.750(-04) &  3.918(-02) & -4.758(+00) &  3.758(+02) & -2.155(+00) &  \\
		&                 &         & ac  & 1.784(-05) &  1.838(-05) & -1.567(-04) & 3.903(-02) &-4.719(+00) &  3.711(+02) &  3.689(+02) & 1.807(-05)\\
				\hline\noalign{\smallskip}
		\multirow{2}{*}{1}  & \multirow{2}{*}{104647.71}       & \multirow{2}{*}{2.81(+01)} & & 3.143(-06) &  5.310(-08) &  6.767(-03) & -4.469(+01) &  1.594(+05) & -3.688(+08) &  6.190(+11) &  \\
		&                 &           & ac & 3.143(-06)& 3.196(-06) &  6.770(-03) & -4.468(+01) & 1.593(+05) & -3.686(+08) &  6.186(+11) & 3.579(-06)\\
		\noalign{\smallskip}\hline \hline
	\end{tabular*}
	\label{tab:Ra_vel}
	    \end{sidewaystable*}

\begin{figure}%
	\centering
	\includegraphics[width=0.75\columnwidth]{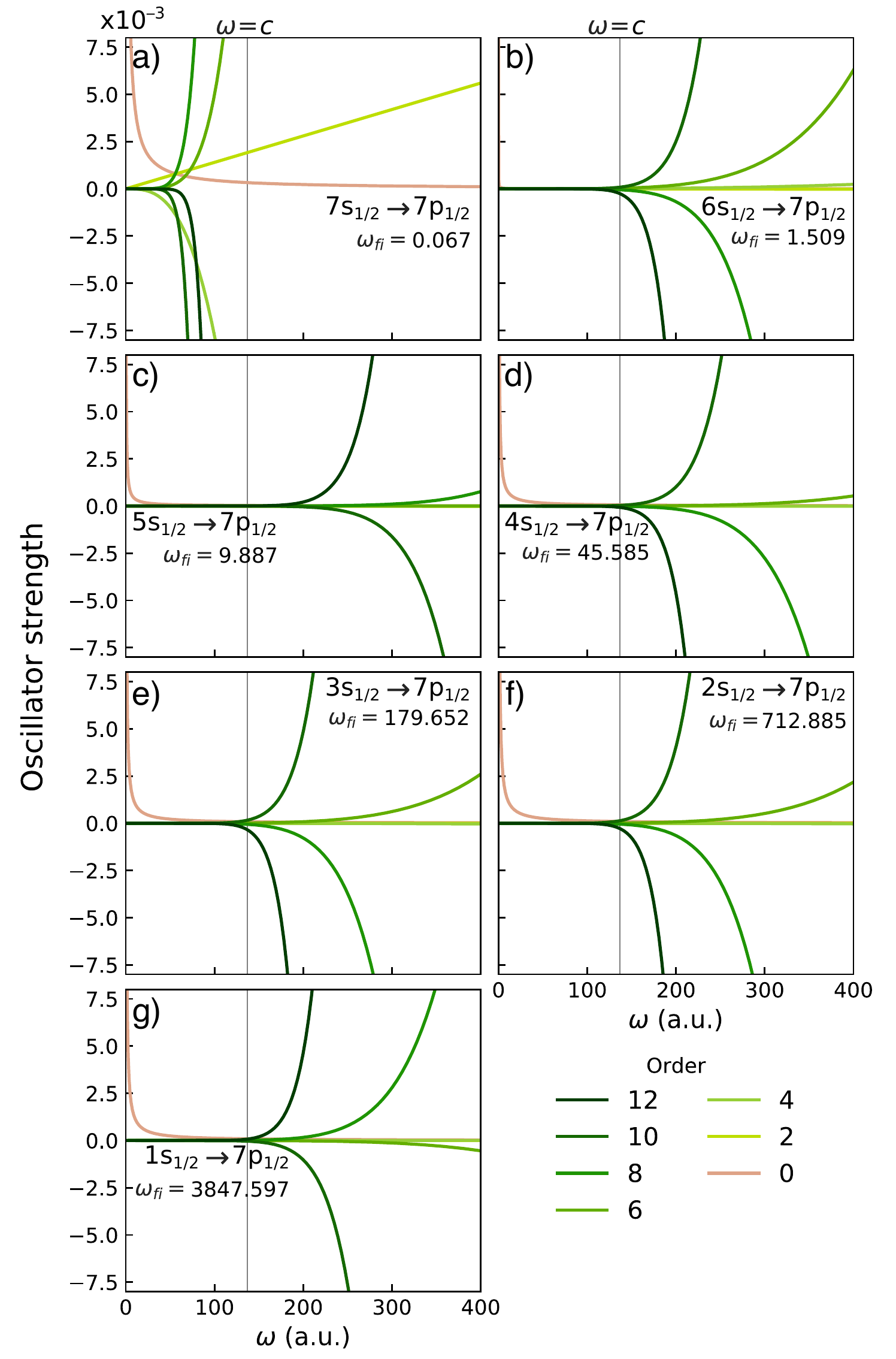}
	\caption{Convergence behavior of the oscillator strengths for $ns_{1/2}\rightarrow7p_{1/2}$ transitions of radium at various orders (colored lines) in the wave vector within the Coulomb gauge (velocity representation) : (a)--(g) correspond to $n=7,6,\dots,1$. Vertical dashed lines indicate $\omega=c$. Excitation energies ($\omega_{fi}$) are in a.u. 
	}
	\label{fig:Ra_convergence_test}%
\end{figure}

We see that for the $7s_{1/2}\rightarrow 7p_{1/2}$ excitation, the electric-dipole approximation holds since the zeroth-order oscillator strength $f^{[0]}$ reproduces the
oscillator strength $f_{\text{full}}$, using the full interaction, to within the reported digits. For other excitations, the second-order oscillator strength $f^{[2]}$ has to be included in order to get reasonable agreement with the full
interaction. For the $1s_{1/2}\rightarrow 7p_{1/2}$ transition, however, higher-order contributions to the oscillator strength blow up. A similar behavior, but to a lesser degree, is observed for the
$2s_{1/2}\rightarrow 7p_{1/2}$ transition, and we also note that the oscillator strength for the $3s_{1/2}\rightarrow 7p_{1/2}$ transition, accumulated to 12th order, is negative. Very similar behavior
is observed for multipolar gauge (data not shown). In Table \ref{tab:Ra_vel} we list for each excitation the corresponding norm $k=\omega/c$ of the wave vector. Interestingly, the apparent divergence in the expansion of the full light-matter interaction occurs when $k\approx 1 \ a_0^{-1}$ (Eq.~\eqref{eq:k}). Indeed, if we do not set $\omega=\omega_{fi}$, where $\hbar\omega_{fi}$ is the excitation energy, and instead treat $\omega$ as a variable, so as to artificially vary $k$ appearing in the interaction operator, we find that the oscillator strengths for all excitations blow up around
$k=1\ a_0^{-1}$, as illustrated in Figure \ref{fig:Ra_convergence_test}. In passing, we note that the excitation energies for Cl $1s\rightarrow$Ti $3d$ transitions of \ce{TiCl4} reported in Table \ref{tab:TICL4} correspond to $k\approx 0.74 \ a_0^{-1}$. It seems reasonable that the convergence behaviour of an expansion of oscillator strengths in orders of the norm of the wave vector should change when $k\approx 1 \ a_0^{-1}$. However, this conclusion requires some caution, since $k$ is not a dimensionless quantity. The proper expansion parameter is rather the dimensionless quantity $kr$ and the above observations suggest that the effective radius $r\approx 1 \ a_0$. For the valence $7s_{1/2}\rightarrow 7p_{1/2}$ excitation the effective radius $r$ is more diffuse, which explains why the apparent divergence sets in for $k < 1 \ a_0^{-1}$, as seen in Figure \ref{fig:Ra_convergence_test}.

The oscillator strengths of given (even) order are calculated according to Eq.~\eqref{eq:osc_vr}. We have also investigated to what extent transition moments over effective interaction operators $\hat{T}_{\mathrm{full}}^{\left[n\right]}$ of order $n$ in the wave vector, Eq.~\eqref{eq:T_n}, sum up to transition moments over the full interaction operator and again find apparent divergences for core excitations. Again, when treating $\omega$ as a variable and not setting it equal to $\omega_{fi}$, we find that these apparent divergences occur for all $ns_{1/2}\rightarrow 7p_{1/2}$ excitations when $k>1\ a_0^{-1}$. Going deeper in our analysis, we note that
transition moments are obtained by contracting the property gradient of the selected operator with the solution vector for the selected excitation, Eq.~\eqref{eq:tmom_general}. Due to the restrictions on the
orbital rotation operator in our particular case, the scalar product is reduced to the multiplication of two numbers. We find that an expansion of the property gradient of the full interaction in orders of the
wave vector displays the same apparent divergence for core excitations as we observed for both oscillator strengths and transition moments. Again, by artificially varying $k$, we
find that these apparent divergences occur when $k>1\ a_0^{-1}$ for \textit{all} excitations.

With our particular orientation of the experiment, the full and truncated effective interaction operator at order $n$ are given by
\begin{equation}\label{eq:Top_case}
\hat{T}_{\mathrm{full}}\left(\omega\right)=\frac{e}{\omega}c\alpha_{z}e^{+iky};\quad\hat{T}_{\mathrm{full}}^{\left[n\right]}(\omega)=\frac{e}{\omega}\frac{i^{n}}{n!}c\alpha_{z}\left(ky\right)^{n}.
\end{equation}
Elements of the property gradient, Eq.~\eqref{eq:property gradient}, of the truncated effective interaction operator are accordingly given by
\begin{equation}\label{eq:pgrad_case}
  g_{T^{[n]};ai} =-\frac{e}{\omega}(ik)^n\left\{\langle\varphi^L_a|c\sigma_z\frac{y^{n}}{n!}|\varphi^S_i\rangle+\langle\varphi^S_a|c\sigma_z\frac{y^{n}}{n!}|\varphi^L_i\rangle
  \right\}
\end{equation}
where superscripts $L$ and $S$ refer to the large and small components of molecular orbital $\varphi_p$, respectively. In practice, as implemented in the \dirac~package, the property gradient is compounded from products of an atomic-orbital (AO) integral with two expansion coefficients on the form
\begin{equation}
  c_{\mu a}^{\ast}\langle\chi_\mu|c\frac{y^{n}}{n!}|\chi_\nu\rangle c_{\nu i},
\end{equation}
with the factor outside the curly brackets in Eq.~\eqref{eq:pgrad_case} multiplied on at the end. In the present case, the coefficients are real due to symmetry.\cite{saue_JCP1999} Each component of the Dirac spinor is expanded in Cartesian Gaussian-type orbitals (CGTOs)
\begin{equation}\label{eq:CGTO}
 G_{ijk}^{\alpha}\left(\boldsymbol{r}\right)=N_{ijk}^{\alpha}x^{i}y^{j}z^{k}e^{-\alpha r^{2}};\quad i+j+k=\ell.
\end{equation}    
For $n=12$ we find that the largest contribution, in terms of magnitude, to the property gradient comes from a small component $p_y$ function with exponent $\alpha_1=1.56556662(-02)\ a_0^{-2}$ combined with a large component $p_y$ function with exponent $\alpha_2=1.24964369(-02)\ a_0^{-2}$. These are the most diffuse $s$ and $p$ functions, respectively, of the large component dyall.ae3z basis set. The resulting AO-integral has a value
$-1.19437467(+6)$ a.u. and is multiplied with a coefficient $c_1=-4.55940113(-8)$ from $1s_{1/2}$ and a coefficient $c_2=-0.844080786$ from $7p_{1/2}$. By calculating AO-integrals with high precision using Mathematica,\cite{Mathematica_11.3} we find that the above AO-integrals, provided by the HERMIT integral package,\cite{prog:hermit} are very stable. On the other hand, the very small $c_1$ coefficient is at the limits of the precision one
can expect from the diagonalization of the Fock matrix, in particular given its ill-conditioning due to the presence of negative-energy solutions. We have, however, investigated the sensitivity of our results with respect to the HF convergence (in terms of the gradient) and find that they are quite stable at tight thresholds.

The final step of our analysis is to study the convergence of the AO-integrals over the truncated interaction towards the corresponding integral over the full interaction operator. Restricting attention to our particular case in Eq.~\eqref{eq:Top_case} and Gaussian $p_y$ functions, in which case only even-order terms contribute, we have 
\begin{equation}
  \langle G_{010}^{\alpha_{1}}|\hat{T}_{\mathrm{full}}(\omega)|G_{010}^{\alpha_{2}}\rangle=\sum_{m=0}^{\infty} \langle G_{010}^{\alpha_{1}}|\hat{T}_{\mathrm{full}}^{\left[2m\right]}(\omega)|G_{010}^{\alpha_{2}}\rangle.
\end{equation}
After eliminating common factors on both sides, we find
an equivalent expression
\begin{equation}\label{eq:Qexpr}
  -(4Q^2-2)e^{-Q^{2}}=\sum_{m=0}^{\infty}\left(-1\right)^{m}a_m;\quad a_m=Q^{2m}\frac{\left(2m+2\right)(2m+1)}{\left(m+1\right)!}
\end{equation}
in terms of a dimensionless parameter $Q$
\begin{align}\label{eq:Q}
Q=\frac{k}{2\sqrt{\alpha_{1}+\alpha_{2}}}
\end{align}
(further details are given in Appendix \ref{sec:AOint}). The right-hand expression has the form of an alternating series and using the Leibniz criterion, we first note  that $\lim\limits_{m\rightarrow\infty}a_m=0$. On the other hand,
the coefficients $a_m$ decrease monotonically only beyond a critical value of the summation index
\begin{equation}
  m_c=\frac{1}{4}\left[\left(2Q^{2}-3\right)+\sqrt{4Q^{4}+12Q^{2}+1}\right].
\end{equation}
For the $1s_{1/2}\rightarrow 7p_{1/2}$ excitation and the above choice of exponents we find that $m_{c}\approx Q^2=6998.7$. For this value of $Q$, the left-hand side of Eq.~\eqref{eq:Qexpr} is essentially zero,
whereas the right-hand side converges extremely slowly towards this value. In fact, using Mathematica,\cite{Mathematica_11.3} no convergence was observed even after summing 10000 terms.
Considering instead the  $2s_{1/2}\rightarrow 7p_{1/2}$ excitation, for which $m_{c}\approx 240$, reasonable convergence is found after summing 282 terms.

In summary we have found that for increasing excitation energies, the use of truncated light-matter interaction becomes increasingly problematic because of the slow convergence of such expansions.
This is not a basis set problem which can be alleviated by increasing the basis set, since we observe this slow convergence at the level of the individual underlying AO-integrals. In particular,
for core excitations, we have observed extremely slow covergence for integrals involving Cartesian Gaussian-type orbitals with diffuse exponents. This can be understood, since such diffuse
functions will be less efficient than tight ones in damping the increasing Cartesian powers appearing in an expansion of the full light-matter interaction in orders of the norm of the wave vector (see Eqs.
\eqref{eq:multipole operator} and \eqref{eq:T_n}). This in turn suggests that the use of Slater-type orbitals, which have slower decay than CGTOs, will be even more problematic. This is indeed the case, as we show in in Appendix \ref{sec:AOint}.

\section{Conclusion}\label{sec:concl}
We have presented the implementation of three schemes for describing light-matter interactions beyond the electric-dipole approximation in the context of linear absorption within the four-component relativistic domain: i) the full semi-classical field--matter interaction operator, in which the electric and magnetic interactions are included to all orders in the wave vector,  in addition to two formulations based on a truncated interaction using either ii) multipolar gauge (generalized length representation) or iii) Coulomb gauge (generalized velocity representation). In the latter gauge,  potentials are given in terms of the values of the electric and magnetic field and their derivatives at some expansion point. We have generalized the derivation of multipolar gauge to arbitrary expansion points and shown that potentials associated with different expansion points are related by a gauge transformation, but also that this is only guaranteed to the extent that the expansion is not truncated. We have further presented schemes for rotational averaging of the oscillator strength for each of the three cases. In particular, the simple form of the light-matter interaction operator in the relativistic formulation allowed for arbitrary-order implementations of the two truncated schemes with and without rotational averaging. We believe that this is a unique feature of our code.

We have next exploited the generality of our formulations and implementation to 
study, both analytically and numerically, the behavior of the two truncated schemes relative to the full light-matter interaction with particular focus on the X-ray spectral region. This analysis has highlighted the following important points:
\begin{itemize}
\item  Oscillator strengths using truncated interaction in Coulomb gauge (generalized velocity representation) are gauge-origin invariant at each order in the wave vector.  This was originally shown in Ref.~\citenum{bernadotte_JCP2012}, but follows straightforwardly from our alternative derivation starting from a Taylor expansion of the full expression for the oscillator strength rather than of the transition moments. A practical realization of this gauge-origin independence, however, relies on an accurate cancellation of terms multiplied by powers of the origin displacement. Thus, while origin invariance is numerically achievable at low frequencies and small displacements, it becomes increasingly difficult, and even unreachable, at higher frequencies and displacements. 
 
\item
Formal gauge-origin invariance of oscillator strengths in multipolar gauge hinges on commutator expressions that do not necessarily hold in a finite basis. 
This explains the notorious lack of order-by-order gauge-origin independence in practical calculations beyond the electric-dipole approximation based on any truncated multipolar gauge formulation.\cite{lestrange_JCP2015,sorensen2017applications} However, we would like to stress that  these commutator relations, involving the
Hamiltonian, correspond to a gauge transformation from the length to the velocity representation. In other words, gauge-origin independence in multipolar gauge is shown by transforming to another gauge for which origin-independence holds. We have not been able to show gauge-origin invariance while staying within
multipolar gauge. An interesting feature of multipolar gauge is that gauge freedom resides within the choice of expansion point. We show that a change of expansion point, that is, gauge origin, corresponds to a gauge transformation, but only if the expansion of the fields is not truncated.
 
\item The appearance of negative oscillator strengths through second order in the wave vector previously reported at the \ce{Cl} \textit{K}-edge for \ce{TiCl4} in the velocity representation\cite{lestrange_JCP2015} is indeed a consequence of a too early truncation of the expansion, as previously suggested.\cite{lestrange_JCP2015,sorensen_CPL2017} In this case, convergence to the full light-matter interaction result is achieved at 12th order in the wave vector irrespective of the basis set used.

\item While the oscillator strengths formulated using truncated interaction in Coulomb gauge (velocity representation) is formally convergent across all frequencies, the series converges extremely slowly at high frequencies, an observation valid also for multipolar gauge.  We report a detailed investigation of a test case where we have studied convergence of the expansion in terms of the wave vector all the way from oscillator strengths to the underlying AO-integrals. For the latter quantities, the expansion in the dimensionless quantity $kr$ is replaced by an expansion in terms of the dimensionless quantity  $Q=k/2\sqrt{\alpha_1+\alpha_2}$, where $\alpha_1$ and $\alpha_2$ are Gaussian exponents. We find that the convergence of integrals over truncated interaction towards integrals over the full interaction is extremely slow, requiring at least $Q^2$ terms. The convergence will depend on the decay of the basis functions. It will be particular slow for diffuse exponents, as can be
  seen from the form of $Q$, and will be worse for Slater-type orbitals than for the  Gaussian-type orbitals used in the present work. 
 The onset of this complication is approximately defined by $\omega= c$ (${\sim}3728$ eV), although it also depends on the size of the given transition moments. Numerical instabilities using Coulomb gauge in the generalized velocity representation can thus be expected already in the higher-energy end of the soft X-ray region even though the onset may be delayed by the order-of-magnitude smaller transition moments associated with core excitations. Caution is therefore necessary using this formulation in simulations of X-ray absorption beyond the electric-dipole approximation because of its practical inapplicability beyond a certain frequency region.
\end{itemize}

The general numerical stability of the full light-matter interaction formulation to gauge-origin transformations and across frequencies as well as its ease of implementation in the context of linear absorption, demonstrated in this work and previously,\cite{List_JCP2015,List_MP2017,sorensen2019implementation} makes this approach the method of choice for simulating linear absorption beyond the electric-dipole approximation. A possible complication of this approach, though, is that the underlying AO-integrals become dependent on the wave vector, hence excitation energies, and must generally be calculated on the fly. 


\begin{acknowledgments}
N.H.L. acknowledges financial support from the Carlsberg Foundation (Grant No.~CF16-0290) and the Villum Foundation (Grant
No.~VKR023371). T.S. would like to thank Anthony Scemama (Toulouse) for the introduction to recursive loops. Computing time from CALMIP (Calcul en Midi-Pyrenées)
and SNIC (Swedish National Infrastructure for Computing) at National
Supercomputer Centre (NSC) are gratefully acknowledged. 

\end{acknowledgments}

\subsection*{Data availability statement}
The data that support the findings of this study are available from the corresponding author upon reasonable request.

\appendix

\section{Simulation of electronic spectra from time-dependent reponse theory\label{sec:Simulation-of-electronic}}

In this Appendix, we provide a brief overview of the simulation of electronic
spectra using time-dependent Hartree--Fock (HF) theory as
implemented in the \dirac~package\cite{DIRAC19} under the restriction
of a closed-shell reference. The formalism carries over with modest
modifications to time-dependent Kohn--Sham (KS) theory.
A fuller account is given in Ref.~\citenum{Bast_IJQC2009} and references
therein. 

We start from a Hamiltonian on the form 
\begin{equation}
\hat{H}=\hat{H}_{0}+\hat{V}\left(t\right);\quad\hat{V}\left(t\right)=\sum_{k=-N}^{+N}\hat{V}\left(\omega_{k}\right)e^{-i\omega_{k}t};\quad\hat{V}\left(\omega_{k}\right)=\sum_{X}\varepsilon_{X}\left(\omega_{k}\right)\hat{h}_{X},
\end{equation}
where appear perturbation strengths $\varepsilon_{X}\left(\omega_{k}\right)$.
All frequencies $\omega_{k}$ are assumed to be integer multiples
of a fundamental frequency $\omega_{T}=2\pi/T$, such that the Hamiltonian
is periodic of period $T$, allowing us to use the quasienergy formalism.
\cite{sambe_PRA1973,christiansen_IJQC1998,saue2002relprop} We employ
a unitary exponential parametrization of the closed-shell HF (or KS)
determinant
\begin{equation}
|\tilde{0}\left(t\right)\rangle=\exp\left[-\hat{\kappa}\left(t\right)\right]|0\rangle
\end{equation}
in terms of an anti-Hermitian, time-dependent orbital rotation operator
\begin{equation}\label{eq:orbrot}
\hat{\kappa}\left(t\right)=\sum_{ai}\left\{ \kappa_{ai}\left(t\right)a_{a}^{\dagger}a_{i}^{\phantom{\dagger}}-\kappa_{ai}^{\ast}\left(t\right)a_{i}^{\dagger}a_{a}^{\phantom{\dagger}}\right\} ;\quad\kappa_{pq}\left(t\right)=\sum_{k=-N}^{+N}\kappa_{pq}\left(\omega_{k}\right)e^{-i\omega_{k}t}.
\end{equation}
Here and in the following indices $\left(i,j\ldots\right)$, $\left(a,b,\ldots\right)$
and $\left(p,q,\ldots\right)$ refer to occupied, virtual and general
orbitals, respectively. 
The linear reponse of the system with respect to some perturbation
$\hat{h}_{B}$ is found from the first-order response equation
\begin{equation}
\left(E_{0}^{[2]}-\hbar\omega_{b}S^{[2]}\right)\boldsymbol{X}_{B}\left(\omega_{b}\right)=-\boldsymbol{E}_{B}^{[1]},\label{eq:first-order response equation}
\end{equation}
where appears the electronic Hessian
\begin{equation}
E_{0}^{[2]}=\left[\begin{array}{cc}
A & B\\
B^{\ast} & A^{\ast}
\end{array}\right];\quad\begin{array}{lcl}
A_{ai,bj} & = & \langle0|\left[-\hat{a}_{i}^{\dagger}\hat{a}_{a}^{\phantom{\dagger}},\left[\hat{a}_{b}^{\dagger}\hat{a}_{j}^{\phantom{\dagger}},\hat{H}_{0}\right]\right]|0\rangle\\[3pt]
B_{ai,bj} & = & \langle0|\left[\hat{a}_{i}^{\dagger}\hat{a}_{a}^{\phantom{\dagger}},\left[\hat{a}_{j}^{\dagger}\hat{a}_{b}^{\phantom{\dagger}},\hat{H}_{0}\right]\right]|0\rangle
\end{array},\label{eq:electronic Hessian}
\end{equation}
the generalized metric
\begin{equation}
S^{[2]}=\left[\begin{array}{cc}
\Sigma & \Delta\\
-\Delta^{\ast} & -\Sigma^{\ast}
\end{array}\right];\quad\begin{array}{lcl}
\Sigma_{ai,bj} & = & \langle0|\left[\hat{a}_{i}^{\dagger}\hat{a}_{a}^{\phantom{\dagger}},\hat{a}_{b}^{\dagger}\hat{a}_{j}^{\phantom{\dagger}}\right]|0\rangle=\delta_{ab}\delta_{ij}\\[3pt]
\Delta_{ai,bj} & = & \langle0|\left[\hat{a}_{i}^{\dagger}\hat{a}_{a}^{\phantom{\dagger}},-\hat{a}_{j}^{\dagger}\hat{a}_{b}^{\phantom{\dagger}}\right]|0\rangle=0
\end{array},\label{eq:generalized metric}
\end{equation}
and the property gradient
\begin{equation}
\boldsymbol{E}_{B}^{[1]}=\left[\begin{array}{c}
\boldsymbol{g}_{B}\\
\Theta_{hB}\boldsymbol{g}_{B}^{\ast}
\end{array}\right];\quad g_{B;ai}=-h_{B;ai}.\label{eq:property gradient}
\end{equation}
An important generalization above is that, in addition to Hermitian
operators $\hat{h}_{B}$ ($\Theta_{hB}=+1$), imposed by the tenets
of quantum mechanics, we also allow anti-Hermitian ones ($\Theta_{hB}=-1$).
It may seem awkward to speak about hermiticity of a vector, but the elements
of the vector are, as seen from Eq.~\eqref{eq:property gradient}, two-index
quantities selected from a matrix and accordingly inherit the symmetries of
that matrix.

The solution vector collects first-order frequency-dependent amplitudes
\begin{equation}
\boldsymbol{X}_{B}\left(\omega_{b}\right)=\left[\begin{array}{c}
\boldsymbol{Z}\\
\boldsymbol{Y}^{\ast}
\end{array}\right];\quad\begin{array}{lcl}
Z_{ai} & = & \left[{\displaystyle \frac{\partial\kappa_{ai}\left(\omega_{b}\right)}{\partial\varepsilon_{B}\left(\omega_{b}\right)}}\right]_{\boldsymbol{\varepsilon}=\boldsymbol{0}}=\kappa_{ai}^{B}\left(\omega_{b}\right)\\
\\
Y_{ai} & = & \left[{\displaystyle \frac{\partial\kappa_{ai}\left(-\omega_{b}\right)}{\partial\varepsilon_{B}\left(\omega_{b}\right)}}\right]_{\boldsymbol{\varepsilon}=\boldsymbol{0}}=\kappa_{ai}^{B}\left(-\omega_{b}\right)
\end{array},\label{eq:solution vectors}
\end{equation}
and linear reponse functions are obtained by contracting solution
vectors with property gradients, that is
\begin{equation}
\langle\langle\hat{A};\hat{B}\rangle\rangle_{\omega_{b}}=\boldsymbol{E}_{A}^{[1]\dagger}\boldsymbol{X}_{B}.
\end{equation}
Excitation energies and corresponding transition moments, on the other
hand, are found from the closely related general eigenvalue problem
\begin{equation}
\left(E_{0}^{[2]}-\hbar\lambda_{m}S^{[2]}\right)\boldsymbol{X}_{m}=0.\label{eq:excitation energy eigenvalue equation}
\end{equation}

From the structure of the electronic Hessian $E_{0}^{\left[2\right]}$, 
Eq.~\eqref{eq:electronic Hessian}, and the general matrix $S^{[2]}$, Eq.~\eqref{eq:generalized metric},
it can be shown that solution vectors of both the first-order response
equation, Eq.~\eqref{eq:first-order response equation}, and the eigenvalue
equation, Eq.~\eqref{eq:excitation energy eigenvalue equation}, come in
pairs
\begin{align}
\lambda_{+;m}&=+\left|\omega_{m}\right|,\,\boldsymbol{X}_{+;m}=\left[\begin{array}{c}
\boldsymbol{Z}_{m}\\
\boldsymbol{Y}_{m}^{\ast}
\end{array}\right]\notag\\
\lambda_{-;m}&=-\left|\omega_{m}\right|,\,\boldsymbol{X}_{-;m}=\left[\begin{array}{c}
\boldsymbol{Y}_{m}\\
\boldsymbol{Z}_{m}^{\ast}
\end{array}\right].
\end{align}
For Hermitian operators $\hat{h}_{A}$ transition moments are obtained
by the contractions
\begin{equation}
\begin{array}{rcccl}
\langle0|\hat{h}_{A}|n\rangle & = & \boldsymbol{E}_{A}^{[1]\dagger}\boldsymbol{X}_{+;n} & = & \boldsymbol{X}_{-;n}^{\dagger}\boldsymbol{E}_{A}^{[1]}\\
\\
\langle n|\hat{h}_{A}|0\rangle & = & \boldsymbol{X}_{+;n}^{\dagger}\boldsymbol{E}_{A}^{[1]} & = & \boldsymbol{E}_{A}^{[1]\dagger}\boldsymbol{X}_{-;n}
\end{array};\quad\Theta_{hA}=+1\label{eq:tmom_H}
\end{equation}

A particular feature of the \dirac~package \cite{DIRAC19} is that
a symmetry scheme, based on quaternion algebra, is applied at the
self-consistent field level and provides automatically maximum point group and time-reversal
symmetry reduction of the computational effort.\cite{saue_JCP1999}
However, the symmetry scheme is restricted to time-symmetric operators
only since their matrix representations in a finite basis can be block
diagonalized by a quaternion unitary transformation.\cite{roesch_CP1983}
In order to accomodate time-antisymmetric, Hermitian operators, they
are made time-symmetric, anti-Hermitian by multiplication with imaginary
$i$,\cite{saue_JCP2003} that is
\begin{equation}
\hat{h}_{A}\rightarrow\hat{h}_{A^{\prime}}=i\hat{h}_{A}\quad\Rightarrow\quad\boldsymbol{E}_{A}^{[1]}\rightarrow\boldsymbol{E}_{A^{\prime}}^{[1]}=i\boldsymbol{E}_{A}^{[1]}.
\end{equation}
For consistency we therefore have to generalize the above relations, Eq.~\eqref{eq:tmom_H}, to
\begin{equation}
\begin{array}{rcccl}
\langle0|\hat{h}_{A}|n\rangle & = & \Theta_{hA}\boldsymbol{E}_{A}^{[1]\dagger}\boldsymbol{X}_{+;n} & = & \boldsymbol{X}_{-;n}^{\dagger}\boldsymbol{E}_{A}^{[1]}\\
\\
\langle n|\hat{h}_{A}|0\rangle & = & \boldsymbol{X}_{+;n}^{\dagger}\boldsymbol{E}_{A}^{[1]} & = & \Theta_{hA}\boldsymbol{E}_{A}^{[1]\dagger}\boldsymbol{X}_{-;n}
\end{array}\label{eq:tmom_general}
\end{equation}

An important observation is that whereas the matrix of time-dependent
amplitudes $\kappa_{pq}\left(t\right)$ is anti-Hermitian, the matrix
of \textit{frequency}-dependent amplitudes $\kappa_{pq}\left(\omega_{k}\right)$,
from which solution vectors are built (cf. Eq. (\ref{eq:solution vectors})),
is general, that is
\begin{equation}
\kappa_{pq}\left(t\right)=-\kappa_{qp}^{\ast}\left(t\right)\quad\Rightarrow\quad\kappa_{qp}^{\ast}\left(-\omega_{k}\right)=-\kappa_{pq}\left(\omega_{k}\right).
\end{equation}
A key to computational efficiency is to consider a decomposition of
solution vectors in terms of components of well-defined hermiticity
and time reversal symmetry.\cite{saue2002relprop,saue_JCP2003} Using a pair of solution vectors $\boldsymbol{X}_{+}$
and $\boldsymbol{X}_{-}$, we may form Hermitian and anti-Hermitian
combinations 
\begin{eqnarray}
	\boldsymbol{X}_{h} & = & \frac{1}{2}\left(\boldsymbol{X}_{+}+\boldsymbol{X}_{-}\right)=\left[\begin{array}{c}
		\boldsymbol{Z}+\boldsymbol{Y}\\
		\boldsymbol{Y}^{\ast}+\boldsymbol{Z}^{\ast}
	\end{array}\right]=\left[\begin{array}{c}
		\boldsymbol{h}\\
		\boldsymbol{h}^{\ast}
	\end{array}\right]\\
	\boldsymbol{X}_{a} & = & \frac{1}{2}\left(\boldsymbol{X}_{+}-\boldsymbol{X}_{-}\right)=\left[\begin{array}{c}
		\boldsymbol{Z}-\boldsymbol{Y}\\
		\boldsymbol{Y}^{\ast}-\boldsymbol{Z}^{\ast}
	\end{array}\right]=\left[\begin{array}{c}
		\boldsymbol{a}\\
		\boldsymbol{a}^{\ast}
	\end{array}\right].
\end{eqnarray}
The inverse relations therefore provide a separation of solution vectors
into Hermitian and anti-Hermitian contributions
\begin{equation}
\boldsymbol{X}_{+}=\boldsymbol{X}_{h}+\boldsymbol{X}_{a};\quad\boldsymbol{X}_{-}=\boldsymbol{X}_{h}-\boldsymbol{X}_{a}.
\end{equation}
Further decomposition of each contribution into time-symmetric and
time-antisymmetric parts gives vectors that are well-defined with respect
to both hermiticity and time reversal symmetry
\begin{equation}
\boldsymbol{U}^{\dagger}\left(\Theta_{h},\Theta_{t}\right)=\left[\begin{array}{cccccccc}
\boldsymbol{c}^{\dagger} & \boldsymbol{d}^{\dagger} & \Theta_{t}\boldsymbol{c}^{T} & \Theta_{t}\boldsymbol{d}^{T} & \Theta_{h}\boldsymbol{c}^{T} & \Theta_{h}\boldsymbol{d}^{T} & \Theta_{h}\Theta_{t}\boldsymbol{c}^{\dagger} & \Theta_{h}\Theta_{t}\boldsymbol{d}^{\dagger}\end{array}\right];\quad\left\{ \begin{array}{ccc}
c_{ai} & = & x_{ai}\\
\\
d_{ai} & = & x_{a\overline{i}}
\end{array}\right.,\label{eq:trial vector}
\end{equation}
where the index overbar refers to a Kramers' partner in a Kramers-restricted
orbital set. The scalar product of such vectors is given by\cite{saue2002relprop}
\begin{equation}
\boldsymbol{U}_{1}^{\dagger}\left(\Theta_{h1},\Theta_{t1}\right)\boldsymbol{U}_{2}\left(\Theta_{h2},\Theta_{t2}\right)=\left(1+\Theta_{h1}\Theta_{h2}\Theta_{t1}\Theta_{t2}\right)\left[z+\Theta_{h1}\Theta_{h2}z^{\ast}\right];\quad z=\left(\boldsymbol{c}_{1}^{\dagger}\boldsymbol{c}_{2}+\boldsymbol{d}_{1}^{\dagger}\boldsymbol{d}_{2}\right),
\end{equation}
and one may therefore distinguish three cases
\begin{equation}
\boldsymbol{U}_{1}^{\dagger}\left(\Theta_{h1},\Theta_{t1}\right)\boldsymbol{U}_{2}\left(\Theta_{h2},\Theta_{t2}\right)=\begin{cases}
0 & ;\quad\Theta_{h1}\Theta_{h2}=-\Theta_{t1}\Theta_{t2}\\
4\text{Re}\left[z\right] & ;\quad\Theta_{h1}\Theta_{h2}=\Theta_{t1}\Theta_{t2}=+1\\
4i\text{Im}\left[z\right] & ;\quad\Theta_{h1}\Theta_{h2}=\Theta_{t1}\Theta_{t2}=-1
\end{cases}.\label{eq:scalar product of trial vectors}
\end{equation}
One may show that hermiticity is conserved when multiplying a vector, Eq.~\eqref{eq:trial vector}, by the electronic Hessian, whereas it is
reversed by the generalized metric. On the other hand, both the electronic Hessian
and the generalized metric conserve time reversal symmetry. The implication
is that the time-symmetric and time-antisymmetric components of a
solution vector do not mix upon solving the generalized eigenvalue
problem, Eq.~\eqref{eq:excitation energy eigenvalue equation}, or the first-order
response equation, Eq.~\eqref{eq:first-order response equation}, and one
can dispense with one of them. From a physical point of view, this
can be understood from the observation that excited states can be
reached through both time-symmetric and time-antisymmetric operators.
From a more practical point of view, this leads to computational savings
corresponding to those obtained by re-expressing the generalized eigenvalue
problem, Eq.~\eqref{eq:excitation energy eigenvalue equation}, as a Hermitian
one of half the dimension and involving the square of transition energies.
Such a transformation can be done exactly in non-relativistic theory,
\cite{Jorgensen_Linderberg_IJQC1970,Jorgensen_annrevphyschem1975,casida:tddftrev1995}
but only through approximations in the relativistic domain.\cite{Peng_Zou_Liu_2005,Wang_Ziegler_Baerends_2005}
In the present scheme, we obtain the same computational savings without
resorting to any transformations or approximations. In order to employ
the quaternion symmetry scheme, we choose to work with the time-symmetric
vectors. It follows from Eq.~\eqref{eq:scalar product of trial vectors}
that their scalar products are either zero or real. In practice, a
property gradient is therefore always contracted with the component
of the solution vector having the same hermiticity, so that all transition
moments are real.

\section{Multipolar gauge\label{sec:Multipolar-gauge}}

In this Appendix, we present a compact derivation of multipolar gauge, following to a large extent Bloch\cite{bloch:gauge} and avoiding indices. We shall write the Taylor expansion of the scalar and vector potential about a reference point $\mathbf{a}$ as
\begin{equation}
\begin{array}{lcl}
\widetilde{\phi}\left(\mathbf{r},t\right) & = & {\displaystyle \sum_{n=0}^{\infty}\frac{1}{n!}\left[\left(\boldsymbol{\delta}_\text{a}\cdot\boldsymbol{\nabla}^{\prime}\right)^{n}\tilde{\phi}\left(\mathbf{r}^{\prime},t\right)\right]_{\mathbf{r}^{\prime}=\mathbf{a}}}\\
\\
\tilde{\mathbf{A}}\left(\mathbf{r},t\right) & = & {\displaystyle \sum_{n=0}^{\infty}\frac{1}{n!}\left[\left(\boldsymbol{\delta}_\text{a}\cdot\boldsymbol{\nabla}^{\prime}\right)^{n}\tilde{\mathbf{A}}\left(\mathbf{r}^{\prime},t\right)\right]_{\mathbf{r}^{\prime}=\mathbf{a}}}
\end{array};\quad\boldsymbol{\delta}_\text{a}=\mathbf{r}-\mathbf{a}
\end{equation}
 We then use the relation $\mathbf{E}=-\boldsymbol{\nabla}\phi-\partial_t\mathbf{A}$
to rewrite the scalar potential as
\begin{equation}
\widetilde{\phi}\left(\mathbf{r},t\right)=\tilde{\phi}\left(\mathbf{a},t\right)-\sum_{n=1}^{\infty}\frac{1}{n!}\left[\left(\boldsymbol{\delta}_\text{a}\cdot\boldsymbol{\nabla}^{\prime}\right)^{n-1}\left(\boldsymbol{\delta}_\text{a}\cdot\mathbf{E}\left(\mathbf{r}^{\prime},t\right)\right)\right]_{\mathbf{r}^{\prime}=\mathbf{a}}-\partial_t\sum_{n=1}^{\infty}\frac{1}{n!}\left[\left(\boldsymbol{\delta}_\text{a}\cdot\boldsymbol{\nabla}^{\prime}\right)^{n-1}\left(\boldsymbol{\delta}_\text{a}\cdot\tilde{\mathbf{A}}\left(\mathbf{r}^{\prime},t\right)\right)\right]_{\mathbf{r}^{\prime}=\mathbf{a}}
\end{equation}
 The scalar potential now has the form of a gauge transformation
\begin{equation}\label{eq:gauge_scalar}
\widetilde{\phi}\left(\mathbf{r},t\right)=\phi\left(\mathbf{r,t}\right)-\partial_t\chi\left(\mathbf{r,t}\right),
\end{equation}
where the gauge function $\chi$ is given by
\begin{equation}\label{eq:mg gauge function}
\chi\left(\mathbf{r},t\right)=\sum_{n=1}^{\infty}\frac{1}{n!}\left[\left(\boldsymbol{\delta}_\text{a}\cdot\boldsymbol{\nabla}^{\prime}\right)^{n-1}\left(\boldsymbol{\delta}_\text{a}\cdot\tilde{\mathbf{A}}\left(\mathbf{r}^{\prime},t\right)\right)\right]_{\mathbf{r}^{\prime}=\mathbf{a}}.
\end{equation}
 Using the partner relation 
\begin{equation}\label{eq:gauge_vector}
\widetilde{\mathbf{A}}\left(\mathbf{r},t\right)=\mathbf{A}\left(\mathbf{r,t}\right)+\boldsymbol{\nabla}\chi\left(\mathbf{r,t}\right),
\end{equation}
 we first work out the gradient of the gauge function to be
\begin{eqnarray}
\boldsymbol{\nabla}\chi\left(\mathbf{r},t\right) & = & \sum_{n=0}^{\infty}\frac{1}{\left(n+1\right)!}\left[\left(\boldsymbol{\delta}_\text{a}\cdot\boldsymbol{\nabla}^{\prime}\right)^{n}\tilde{\mathbf{A}}\left(\mathbf{r}^{\prime},t\right)\right]_{\mathbf{r}^{\prime}=\mathbf{a}}+\sum_{n=1}^{\infty}\frac{n}{\left(n+1\right)!}\left[\left(\boldsymbol{\delta}_\text{a}\cdot\boldsymbol{\nabla}^{\prime}\right)^{n-1}\boldsymbol{\nabla}^{\prime}\left(\boldsymbol{\delta}_\text{a}\cdot\tilde{\mathbf{A}}\left(\mathbf{r}^{\prime},t\right)\right)\right]_{\mathbf{r}^{\prime}=\mathbf{a}}.
\end{eqnarray}
 Further manipulation then gives
\begin{eqnarray}
\mathbf{A}\left(\mathbf{r,t}\right) & = & \sum_{n=1}^{\infty}\frac{n}{\left(n+1\right)!}\left[\left(\boldsymbol{\delta}_\text{a}\cdot\boldsymbol{\nabla}^{\prime}\right)^{n-1}\left\{ \left(\boldsymbol{\delta}_\text{a}\cdot\boldsymbol{\nabla}^{\prime}\right)\tilde{\mathbf{A}}\left(\mathbf{r}^{\prime},t\right)-\boldsymbol{\nabla}^{\prime}\left(\boldsymbol{\delta}_\text{a}\cdot\tilde{\mathbf{A}}\left(\mathbf{r}^{\prime},t\right)\right)\right\} \right]_{\mathbf{r}^{\prime}=\mathbf{a}}.
\end{eqnarray}
 Finally, using the relation
\begin{equation}\label{eq:deltaxB}
\boldsymbol{\delta}\times\mathbf{B}=\boldsymbol{\delta}\times\left(\boldsymbol{\nabla}\times\tilde{\mathbf{A}}\right)=\boldsymbol{\nabla}\left(\boldsymbol{\delta}\cdot\tilde{\mathbf{A}}\right)-\left(\boldsymbol{\delta}\cdot\boldsymbol{\nabla}\right)\tilde{\mathbf{A}},
\end{equation}
 we arrive at the final form of the potentials
 \begin{equation}\label{eq:mg potentials}
   \begin{array}{lcl}
  \phi_\text{a}\left(\mathbf{r},t\right) & = & {\displaystyle \tilde{\phi}\left(\mathbf{a},t\right)-\sum_{n=0}^{\infty}\frac{1}{\left(n+1\right)!}\left[\left(\boldsymbol{\delta}_\text{a}\cdot\boldsymbol{\nabla}^{\prime}\right)^{n}\left(\boldsymbol{\delta}_\text{a}\cdot\mathbf{E}\left(\mathbf{r}^{\prime},t\right)\right)\right]_{\mathbf{r}^{\prime}=\mathbf{a}}}\\
  \mathbf{A}_\text{a}\left(\mathbf{r,t}\right) & = & {\displaystyle -\sum_{n=1}^{\infty}\frac{n}{\left(n+1\right)!}\left[\left(\boldsymbol{\delta}_\text{a}\cdot\boldsymbol{\nabla}^{\prime}\right)^{n-1}\left(\boldsymbol{\delta}_\text{a}\times\mathbf{B}\left(\mathbf{r}^{\prime},t\right)\right)\right]_{\mathbf{r}^{\prime}=\mathbf{a}}}
  \end{array}
\end{equation}

An alternative derivation of multipolar gauge,\cite{Kobe_AJP1980,Brittin_AJP1982,Kobe_AJP1983,Kobe_INC1985} that we here generalize to an arbitrary expansion point,
is obtained by integrating Eq.~\eqref{eq:gauge_vector}
along a line from expansion point $\mathbf{a}$ to observer point
$\boldsymbol{r}$ and setting the gauge condition
\begin{equation}
\boldsymbol{\delta}_{\text{a}}\cdot\int_{0}^{1}\mathbf{A}\left(\lambda\boldsymbol{\delta}_{\text{a}}+\mathbf{a},t\right)d\lambda=0.
\end{equation}
 The gauge function is then found to be
\begin{equation}
\chi_{\text{a}}\left(\mathbf{r,t}\right)=\int_{0}^{1}\boldsymbol{\delta}_{\text{a}}\cdot\tilde{\mathbf{A}}\left(\lambda\boldsymbol{\delta}_{\text{a}}+\mathbf{a},t\right)d\lambda,
\end{equation}
and the resulting potentials read
\begin{equation}\label{eq:mg potentials2}
\begin{array}{lcl}  
\phi_{\text{a}}\left(\mathbf{r},t\right) & = & \tilde{\phi}\left(\mathbf{a},t\right)-\boldsymbol{\delta}_{\text{a}}\cdot\int_{0}^{1}\mathbf{E}\left(\lambda\boldsymbol{\delta}_{\text{a}}+\mathbf{a},t\right)d\lambda\\
\mathbf{A}_{\text{a}}\left(\mathbf{r,t}\right) & = & -\int_{0}^{1}\lambda\left[\boldsymbol{\delta}_{\text{a}}\times\mathbf{B}\left(\lambda\boldsymbol{\delta}_{\text{a}}+\mathbf{a},t\right)\right]d\lambda,
\end{array}
\end{equation}
where we have used Eq.~\eqref{eq:deltaxB}. The equivalence of the expressions of the present paragraph with those of the preceding one is seen by expanding the functions of $\mathbf{r}^{\prime}=\lambda\boldsymbol{\delta}_{\mathbf{a}}+\mathbf{a}$ in the integrands about
$\mathbf{r}^{\prime}=\mathbf{a}$.

In passing we note that the divergence of the vector potential is given by
\begin{eqnarray}
\boldsymbol{\nabla}\cdot\mathbf{A}\left(\mathbf{r,t}\right)
 & = & \int_{0}^{1}\lambda\boldsymbol{\delta}_{\text{a}}\cdot\left(\boldsymbol{\nabla}\times\mathbf{B}\left(\lambda\boldsymbol{\delta}_{\text{a}}+\mathbf{a},t\right)\right)d\lambda\\
 & = & \int_{0}^{1}\lambda^{2}\boldsymbol{\delta}_{\text{a}}\cdot\left(\mu_{0}\mathbf{j}\left(\lambda\boldsymbol{\delta}_{\text{a}}+\mathbf{a},t\right)+\frac{1}{c^{2}}\partial_{t}\mathbf{E}\left(\lambda\boldsymbol{\delta}_{\text{a}}+\mathbf{a},t\right)\right)d\lambda,
\end{eqnarray}
where appears the magnetic constant $\mu_{0}$ and the current
density $\mathbf{j}$. The Amp{\`e}re--Maxwell law was used in the final step. 
This relation shows that the multipolar gauge is
equivalent to the Coulomb gauge only in the absence of external currents
and for static electric fields.

In multipolar gauge the potentials are given in terms of the fields and their derivatives at the selected expansion point, which seems to eliminate any gauge freedom.
However, this is incorrect. The gauge freedom is retained in the free choice of the expansion point. Consider now the gauge transformation taking us from potentials
$(\mathbf{A}_\text{a},\phi_\text{a})$, defined with respect to expansion point $\mathbf{a}$, to a new set of potentials $(\mathbf{A}_\text{b},\phi_\text{b})$, defined with respect to expansion point $\mathbf{b}$. Clearly the gauge function $\chi_{\text{a}\rightarrow \text{b}}$ satisfies
\begin{eqnarray}
  \partial_t\chi_{\text{a}\rightarrow \text{b}}\left(\mathbf{r,t}\right) & = & \phi_{\text{a}}\left(\mathbf{r},t\right)-\phi_{\text{b}}\left(\mathbf{r,t}\right)\label{eq:scalar_ab}\\
  \boldsymbol{\nabla}\chi_{\text{a}\rightarrow \text{b}}\left(\mathbf{r,t}\right) & = & \mathbf{A}_{\text{b}}\left(\mathbf{r},t\right)-\mathbf{A}_{\text{a}}\left(\mathbf{r},t\right)\label{eq:vector_ab}
\end{eqnarray}
Starting from Eq.~\eqref{eq:scalar_ab} we find that
{\small\begin{eqnarray}
\boldsymbol{\nabla}\partial_t\chi_{\text{a}\rightarrow \text{b}}\left(\mathbf{r,t}\right) & = & \sum_{n=1}^{\infty}\frac{n}{\left(n+1\right)!}\left\{ \left[\left(\boldsymbol{\delta}_{\text{b}}\cdot\boldsymbol{\nabla}^{\prime}\right)^{n-1}\boldsymbol{\nabla}^{\prime}\left(\boldsymbol{\delta}_{\text{b}}\cdot\mathbf{E}\left(\mathbf{r}^{\prime},t\right)\right)\right]_{\mathbf{r}^{\prime}=\mathbf{b}}-\left[\left(\boldsymbol{\delta}_{\text{a}}\cdot\boldsymbol{\nabla}^{\prime}\right)^{n-1}\boldsymbol{\nabla}^{\prime}\left(\boldsymbol{\delta}_{\text{a}}\cdot\mathbf{E}\left(\mathbf{r}^{\prime},t\right)\right)\right]_{\mathbf{r}^{\prime}=\mathbf{a}}\right\} \notag\\
 & + & \sum_{n=0}^{\infty}\frac{1}{\left(n+1\right)!}\left\{ \left[\left(\boldsymbol{\delta}_{\text{b}}\cdot\boldsymbol{\nabla}^{\prime}\right)^{n}\mathbf{E}\left(\mathbf{r}^{\prime},t\right)\right]_{\mathbf{r}^{\prime}=\mathbf{b}}-\left[\left(\boldsymbol{\delta}_{\text{a}}\cdot\boldsymbol{\nabla}^{\prime}\right)^{n}\mathbf{E}\left(\mathbf{r}^{\prime},t\right)\right]_{\mathbf{r}^{\prime}=\mathbf{a}}\right\} 
\end{eqnarray}}
On the other hand, starting from Eq.~\eqref{eq:vector_ab}, we find that
{\small{}
\begin{eqnarray*}
\partial_t\boldsymbol{\nabla}\chi_{\text{a}\rightarrow \text{b}}\left(\mathbf{r,t}\right) 
 & = & \sum_{n=1}^{\infty}\frac{n}{\left(n+1\right)!}\left\{ \left[\left(\boldsymbol{\delta}_{\text{b}}\cdot\boldsymbol{\nabla}^{\prime}\right)^{n-1}\boldsymbol{\nabla}^{\prime}\left(\boldsymbol{\delta}_{\text{b}}\cdot\mathbf{E}\left(\mathbf{r}^{\prime},t\right)\right)\right]_{\mathbf{r}^{\prime}=\mathbf{b}}-\left[\left(\boldsymbol{\delta}_{\text{a}}\cdot\boldsymbol{\nabla}^{\prime}\right)^{n-1}\boldsymbol{\nabla}^{\prime}\left(\boldsymbol{\delta}_{\text{a}}\cdot\mathbf{E}\left(\mathbf{r}^{\prime},t\right)\right)\right]_{\mathbf{r}^{\prime}=\mathbf{a}}\right\} \\
 & - & \sum_{n=1}^{\infty}\frac{n}{\left(n+1\right)!}\left\{ \left[\left(\boldsymbol{\delta}_{\text{b}}\cdot\boldsymbol{\nabla}^{\prime}\right)^{n}\mathbf{E}\left(\mathbf{r}^{\prime},t\right)\right]_{\mathbf{r}^{\prime}=\mathbf{b}}-\left[\left(\boldsymbol{\delta}_{\text{a}}\cdot\boldsymbol{\nabla}^{\prime}\right)^{n}\mathbf{E}\left(\mathbf{r}^{\prime},t\right)\right]_{\mathbf{r}^{\prime}=\mathbf{a}}\right\} ,
\end{eqnarray*}
}
where we have used Faraday's law
\begin{equation}
  \boldsymbol{\nabla}\times\mathbf{E}+\partial_{t}\mathbf{B}=0.
\end{equation}
Due to commutation of space and time derivatives the two expressions
should be the same, provided that the potentials at the two expansion points are
related by a gauge transformation. At first sight, this does not seem to be the case, since the second line of the above expressions differ. However, actually calculating the difference gives
\begin{eqnarray}
\boldsymbol{\nabla}\partial_t\chi_{\text{a}\rightarrow \text{b}}\left(\mathbf{r,t}\right)-\partial_t\boldsymbol{\nabla}\chi_{\text{a}\rightarrow \text{b}}\left(\mathbf{r,t}\right) 
 & = & \sum_{n=0}^{\infty}\frac{1}{n!}\left\{ \left[\left(\boldsymbol{\delta}_{\text{b}}\cdot\boldsymbol{\nabla}^{\prime}\right)^{n}\mathbf{E}\left(\mathbf{r}^{\prime},t\right)\right]_{\mathbf{r}^{\prime}=\mathbf{b}}-\left[\left(\boldsymbol{\delta}_{\text{a}}\cdot\boldsymbol{\nabla}^{\prime}\right)^{n}\mathbf{E}\left(\mathbf{r}^{\prime},t\right)\right]_{\mathbf{r}^{\prime}=\mathbf{a}}\right\} =0,
\end{eqnarray}
which is zero since the final line is the difference of the
Taylor expansions of $\mathbf{E}\left(\mathbf{r},t\right)$ at the two different expansion points. However,
a very important observation is that this cancellation, and hence gauge freedom, is \textit{only} assured if the expansions are \textit{not} truncated.

Before closing this brief overview of multipolar gauge, we remark that in some sources a distinction is made between minimal coupling and multipolar Hamiltonians.\cite{Barron_Gray_JPA1973,Power_PL1982,bandrauk_MolLas,Chernyak_CP1995,Salam_PhysRevA1997,Ishikawa_PRA2018} This terminology arises from the observation that gauge transformations in quantum mechanics (and beyond) may be induced by a local unitary transformation of the wave function\cite{Fock_ZP1926,Fock_coll2019,Jackson_RMP2001}
\begin{equation}
  \psi(\mathbf{r},t)\rightarrow\psi^\prime(\mathbf{r},t)=U(\mathbf{r},t)\psi(\mathbf{r},t);\quad U(\mathbf{r},t)=e^{-\frac{i}{\hbar}q\chi(\mathbf{r},t)},
\end{equation}
where appears particle charge $q$, with the corresponding time-dependent wave equation
\begin{equation}
  \left(\hat{\cal H}(\mathbf{A},\phi)-i\hbar\partial_t\right)\psi(\mathbf{r},t)=0\quad\rightarrow\quad
  \left(\hat{\cal H}^\prime(\mathbf{A}^\prime,\phi^\prime)-i\hbar\partial_t\right)\psi^\prime(\mathbf{r},t)=0
\end{equation}
expressed in terms of a transformed Hamiltonian
\begin{equation}\label{eq:Hgauge}
  \hat{\cal H}^\prime=U\hat{\cal H}U^{-1}-i\hbar U\partial_t(U^{-1})=U\hat{\cal H}U^{-1}+q\partial_{t}\chi
\end{equation}
with potentials
\begin{equation}
  \mathbf{A}^{\prime}=\mathbf{A}-\boldsymbol{\nabla}\chi;\quad\phi^{\prime}=\phi+\partial_{t}\chi.
\end{equation}
Accordingly, the multipolar or Power--Zienau--Woolley Hamiltonian\cite{Power_Zienau_PhilTran1959,Atkins_Woolley_RSPA1970,Woolley_RSPA1971,Rousseau_SciRep2017,David_JCP2018} is obtained from transforming the non-relativistic minimal coupling Hamiltonian by using the multipolar gauge function, Eq.~\eqref{eq:mg gauge function}. However, this is possibly misleading terminology since minimal coupling is a general procedure for coupling particles to fields,\cite{gell-mann:1956,saue2002relprop} and, indeed, the multipolar Hamiltonian can equivalently be obtained by plugging in the multipolar gauge potentials,
Eq.\eqref{eq:mg potentials}, into the free-particle Hamiltonian according to the principle of minimal
electromagnetic coupling.\cite{Babiker_Loudon_rspa1983,Chernyak_CP1995}

A final observation is that the transformed Hamiltonian, Eq.\eqref{eq:Hgauge}, using the Baker--Campbell--Hausdorff (BCH) expansion, can alternatively be expressed as a sequence of increasingly nested commutators involving the gauge function and the original Hamiltonian
\begin{equation}
  \hat{\cal H}^{\prime}=\hat{\cal H}+q\partial_{t}\chi-q\frac{i}{\hbar}\left[\chi,\hat{\cal H}\right]-\frac{q^{2}}{2\hbar^{2}}\left[\chi,\left[\chi,\hat{\cal H}\right]\right]+\ldots .
\end{equation}
An illuminating example is to start from the gauge function associated with multipolar gauge, Eq.~\eqref{eq:mg gauge function}. If we introduce the
potentials, Eq.~\eqref{eq:linear plane wave potentials}, associated with
linearly polarized monochromatic light in Coulomb gauge, the gauge function for expansion point $\boldsymbol{a}=\boldsymbol{0}$ can be expressed as
\begin{equation}\label{eq:chicas}
  \chi\left(\mathbf{r},t\right)=
  \chi\left(\mathbf{r},\omega\right)e^{-i\omega t}
  +\chi\left(\mathbf{r},-\omega\right)e^{+i\omega t};\quad
   \chi\left(\mathbf{r},\omega\right)= -\frac{E_{\omega}}{2\omega}\sum_{n=0}^{\infty}\frac{1}{\left(n+1\right)!}\left(i\mathbf{k}\cdot\mathbf{r}\right)^{n}\left(\mathbf{r}\cdot\boldsymbol{\epsilon}\right)e^{i\delta}.
\end{equation}
Using Eq.~\eqref{eq:Tcomm}, we find
\begin{align}\label{eq:chicomm}
  -\frac{i}{\hbar}\left[\chi\left(\mathbf{r},\omega\right),\hat{\cal H}\right]&=
  -\frac{E_{\omega}}{2\omega}\sum_{n=0}^{\infty}\frac{1}{\left(n+1\right)!}\left(c\boldsymbol{\alpha}\cdot\boldsymbol{\epsilon}\right)\left(i\mathbf{k}\cdot\mathbf{r}\right)^{n}e^{i\delta}\notag\\
  &-\frac{E_{\omega}}{2\omega}\sum_{n=0}^{\infty}\frac{n}{\left(n+1\right)!}\left(\mathbf{r}\cdot\boldsymbol{\epsilon}\right)\left(i\mathbf{k}\cdot c\boldsymbol{\alpha}\right)\left(i\mathbf{k}\cdot\mathbf{r}\right)^{n-1}e^{i\delta},
\end{align}
whereas $\left[\chi,\left[\chi,\hat{\cal H}\right]\right]$ and all higher-order commutators in the BCH expansion vanish, as can be seen from Eqs.~\eqref{eq:chicas} and \eqref{eq:chicomm}.

Starting from the light-matter interaction operator, Eq.~\eqref{eq:effT}, in Coulomb gauge
\begin{equation}
\hat{V}_{\mathrm{full}}\left(\mathbf{r},\omega\right)=-e\frac{E_{\omega}}{2\omega}\left(c\boldsymbol{\alpha}\cdot\boldsymbol{\epsilon}\right)e^{i\left(\mathbf{k}\cdot\mathbf{r}+\delta\right)},
\end{equation}
we find that the transformed operator reads
\begin{equation}
\hat{V}^{\prime}\left(\mathbf{r},\omega\right)  =  \hat{V}_{\mathrm{full}}\left(\mathbf{r},\omega\right)+ie\omega\chi\left(\mathbf{r},\omega\right)+\frac{ie}{\hbar}\left[\chi\left(\mathbf{r},\omega\right),\hat{\cal H}\right]=\frac{1}{2}iE_{\omega}\hat{T}_{\mathrm{mg}}e^{i\delta}=-\frac{1}{2}E_{\omega}\hat{T}_{\mathrm{mg}},
\end{equation}
where we have used Eq.~\eqref{eq:useful} and recognize the effective interaction in multipolar gauge, Eq.~\eqref{eq:T_mp}.
The final form is obtained by setting the phase $\delta=\pi/2$ in accordance with the phase convention Eq.~\eqref{eq:phase convention}. 
This derivation thereby demonstrates that the change from velocity to length representation, Eq.~\eqref{eq:ED_v2l} and its generalization in
Eq.~\eqref{eq:BED_v2l} is obtained by a gauge transformation.

\section{The trivariate beta function\label{sec:The-trivariate-beta}}

In this Appendix, we demonstrate the integral representation, Eq.~\eqref{eq:trivariate beta function}, of the trivariate beta function. We start from the integral representation
of the gamma function\cite{Arfken_MatMetPhys2013}
\begin{equation}
\Gamma\left(a\right)=2\int_{0}^{\infty}e^{-x^{2}}x^{2a-1}dx,
\end{equation}
and consider the triple product
\begin{equation}
\Gamma\left(a\right)\Gamma\left(b\right)\Gamma\left(c\right)=8\int_{0}^{\infty}\int_{0}^{\infty}\int_{0}^{\infty}e^{-\left(x^{2}+y^{2}+z^{2}\right)}x^{2a-1}y^{2b-1}z^{2c-1}dxdydz.
\end{equation}
 Noting that the integration is limited to the $(+,+,+)$ octant of
Euclidean space, we switch to spherical coordinates
\begin{eqnarray}
\Gamma\left(a\right)\Gamma\left(b\right)\Gamma\left(c\right) & = & 8\int_{0}^{\pi/2}\int_{0}^{\pi/2}\int_{0}^{\infty}e^{-r^{2}}r^{2(a+b+c)-3}e_{r;x}^{2a-1}e_{r;y}^{2b-1}e_{r:z}^{2c-1}r^{2}\sin\theta drd\theta d\phi\\
 & = & 4\Gamma\left(a+b+c\right)\int_{0}^{\pi/2}\int_{0}^{\pi/2}e_{r;x}^{2a-1}e_{r;y}^{2b-1}e_{r;z}^{2c-1}\sin\theta d\theta d\phi,
\end{eqnarray}
 which leads directly to the introduction of the trivariate beta function
and its integral representation as given in Eq.~\eqref{eq:trivariate beta function}.

\section{Specific integrals over full and truncated light-matter interaction}\label{sec:AOint}
In the test case analyzed in Section \ref{subsec:Ra_truncated}, the wave $\mathbf{k}$ and polarization $\boldsymbol{\epsilon}$ vectors are oriented along the $y$- and $z$-axes, respectively, such that
the full and truncated effective interaction operator at order $n$ are given by Eq.~\eqref{eq:Top_case}.  We want to study the convergence of the underlying AO-integrals over
the truncated interaction towards the corresponding AO-integral over
the full interaction. These involve only the scalar parts of the operators,
so in practice we study the expression
\begin{equation}
\langle\chi_{\mu}|e^{+iky}|\chi_{\nu}\rangle=\sum_{n=0}^{\infty}\frac{\left(ik\right)^{n}}{n!}\langle\chi_{\mu}|y^{n}|\chi_{\nu}\rangle,\label{eq:AO-integrals}
\end{equation}
where $\chi_{\mu}$ and $\chi_{\nu}$ are scalar basis functions.
We shall limit attention to $p_{y}$ functions since the largest
integrals in our study involved such basis functions with diffuse
exponents.

The calculations presented in this paper are based on \emph{Cartesian
Gaussian-type orbitals} (CGTOs), Eq.~\eqref{eq:CGTO}.  With these basis functions the volume integrals on both sides of
Eq.~\eqref{eq:AO-integrals} factorize into integrals over the three
Cartesian components. After elimination of common factors, Eq.~\eqref{eq:AO-integrals}
reduces to
\begin{equation}
\langle G_{j_{1}}^{\alpha_{1}}|e^{+iky}|G_{j_{2}}^{\alpha_{2}}\rangle_{y}=\sum_{n=0}^{\infty}\frac{\left(ik\right)^{n}}{n!}\langle G_{j_{1}}^{\alpha_{1}}|y^{n}|G_{j_{2}}^{\alpha_{2}}\rangle_{y};\quad G_{j}^{\alpha}=N_{j}^{\alpha}y^{j}e^{-\alpha y^{2}}.
\end{equation}
 The left-hand side integral corresponds to a Fourier transform. To
evaluate the integral, we use the formula\cite{strichartz_distributions}
\begin{equation}
{\cal F}\left[e^{-\alpha y^{2}}\right]\left(k\right)=\int_{-\infty}^{+\infty}e^{-\alpha y^{2}}e^{iky}dy=\sqrt{\frac{\pi}{\alpha}}e^{-k^{2}/4\alpha},
\end{equation}
as well as
\begin{equation}
\left(-i\partial_{k}\right)^{n}{\cal F}\left[f(y)\right]\left(k\right)=\int_{-\infty}^{+\infty}y^{n}f\left(y\right)e^{iky}dy,
\end{equation}
to obtain
\begin{equation}
\langle G_{j_{1}}^{\alpha_{1}}|e^{+iky}|G_{j_{2}}^{\alpha_{2}}\rangle_{y}=N_{j_{1}}^{\alpha_{1}}N_{j_{2}}^{\alpha_{2}}\sqrt{\frac{\pi}{\alpha_{1}+\alpha_{2}}}\left(\frac{i}{2\sqrt{\alpha_{1}+\alpha_{2}}}\right)^{j_{1}+j_{2}}e^{-Q^2}H_{\left(j_{1}+j_{2}\right)}\left(Q\right),
\end{equation}
in terms of Hermite polynomials $H_{j}$ and the dimensionless parameter $Q$, Eq.~\eqref{eq:Q}. For the right-hand side
integral we obtain
\begin{equation}
\langle G_{j_{1}}^{\alpha_{1}}|y^{n}|G_{j_{2}}^{\alpha_{2}}\rangle_{y}=N_{j_{1}}^{\alpha_{1}}N_{j_{2}}^{\alpha_{2}}\frac{1}{2}\left[1+\left(-1\right)^{\left(j_{1}+j_{2}+n\right)}\right]\left(\alpha_{1}+\alpha_{2}\right)^{-(j_{1}+j_{2}+n+1)/2}\Gamma\left(\frac{j_{1}+j_{2}+n+1}{2}\right),\label{eq:Gtrunc}
\end{equation}
where we have used the integral representation of the gamma function
\begin{equation}
\Gamma\left(a\right)=2\int_{0}^{\infty}x^{2a-1}e^{-x^{2}}dx.
\end{equation}
In our particular case, we have $j_{1}=j_{2}=1$, and so one sees
from the expression in square brackets of Eq.~\eqref{eq:Gtrunc} that
only even $n=2m$ contributions will be non-zero. Again eliminating
common factors we arrive at Eq.~\eqref{eq:Qexpr}.

Further insight is provided by comparing with corresponding integrals
obtained with \emph{Slater-type orbitals (STOs)}. We shall again limit
attention to $p_{y}$ functions which we express as 
\begin{equation}
S_{y}^{\zeta}=N_{\zeta}y\exp\left[-\zeta r\right].
\end{equation}
 In this case, factorization of integrals over Cartesian components
is no longer possible. For the full interaction we get 
\begin{equation}
\langle S_{y}^{\zeta_{1}}|e^{+iky}|S_{y}^{\zeta_{2}}\rangle_{y}=32\pi N_{\zeta_{1}}N_{\zeta_{2}}\bar{\zeta}\left(\bar{\zeta}^{2}-5k^{2}\right)\left[\bar{\zeta}^{2}+k^{2}\right]^{-4};\quad\bar{\zeta}=\zeta_{1}+\zeta_{2}
\end{equation}
where we have used the Fourier transform\cite{strichartz_distributions}
\begin{equation}
{\cal F}\left[e^{-\zeta r}\right]\left(\boldsymbol{k}\right)=8\pi\frac{\zeta}{\left(\zeta^{2}+k^{2}\right)^{2}}.
\end{equation}
For the truncated interaction, we obtain
\begin{equation}
\langle S_{y}^{\zeta_{1}}|y^{n}|S_{y}^{\zeta_{2}}\rangle=\frac{2\pi}{n+3}\left[1-\left(-1\right)^{n+3}\right]N_{\zeta_{1}}N_{\zeta_{2}}\int_{0}^{\infty}r^{n+4}\exp\left[-\bar{\zeta}r\right]dr,
\end{equation}
where the expression in square brackets, coming from angular integration,
again shows that only even $n=2m$ contributions will be non-zero.
The radial integral is found as
\begin{equation}
\int_{0}^{\infty}r^{2m+4}\exp\left[-\bar{\zeta}r\right]dr=\partial_{\bar{\zeta}}^{2m+4}\int_{0}^{\infty}\exp\left[-\bar{\zeta}r\right]dr=\left(2m+4\right)!\bar{\zeta}^{-(2m+5)}.
\end{equation}
 After elimination of common factors, Eq.~\eqref{eq:AO-integrals} may in this case be expressed as
\begin{equation}\label{eq:STOconv}
8\left(1-5\tilde{Q}^{2}\right)\left[1+\tilde{Q}^{2}\right]^{-4}=\sum_{m=0}^{\infty}\left(-1\right)^{m}\widetilde{a}_{m};\quad\widetilde{a}_{m}=\tilde{Q}^{2m}\left(2m+4\right)\left(2m+2\right)\left(2m+1\right),
\end{equation}
 where we have introduced the dimensionless variable 
\begin{equation}
\tilde{Q}=\frac{k}{\zeta_{1}+\zeta_{2}}.
\end{equation}
 As in the case of CGTOs, the right-hand side has the form of an alternating
series, but now convergence becomes even more problematic since $\lim\limits_{m\rightarrow\infty}\tilde{a}_m=\infty$.
We also note the limit
\begin{equation}
\lim_{m\rightarrow\infty}\frac{\tilde{a}_{m+1}}{\tilde{a}_{m}}=\tilde{Q}^{2}\lim_{m\rightarrow\infty}\frac{\left(m+3\right)\left(2m+3\right)}{\left(m+1\right)\left(2m+1\right)}=\tilde{Q}^{2},
\end{equation}
which is zero in the case of CGTOs. Numerically, we only find convergence of the right-hand side of Eq.~\eqref{eq:STOconv} for $\tilde{Q}<1$.

\bibliography{relbed}

\end{document}